\documentclass[10pt,a4paper,superscriptaddress,aps,prd,showkeys,showpacs,nofootinbib,reprint]{revtex4-1}
\usepackage[utf8]{inputenc}
\usepackage{verbatim}
\usepackage{amsmath,amsfonts,amssymb}
\usepackage[breaklinks,colorlinks]{hyperref}
\usepackage{natbib}
\usepackage{tikz}
\usepackage{float}
\usepackage{graphicx}
\usepackage{adjustbox}
\usepackage{tabularx}
\usepackage{amsbsy}
\usepackage{braket}
\hypersetup{%
,urlcolor=blue
,citecolor=blue
,linkcolor=blue
}

\def\aap{Astronomy and Astrophysics}

\def\jcap{Journal of Cosmology and Astro-Particle Physics}

\begin{document}

\title{Radio Line Properties of Axion Dark Matter Conversion in Neutron Stars}

\author{R.~A.~Battye}
\email[]{richard.battye@manchester.ac.uk}
\affiliation{%
Jodrell Bank Centre for Astrophysics, School of Natural Sciences, Department of Physics and Astronomy, University of Manchester, Manchester, M13 9PL, U.K.
}

\author{B.~Garbrecht}
\email[]{garbrecht@tum.de}
\affiliation{%
Technische Universität München, Physik-Department, James-Franck-Straße, 85748 Garching, Germany
}

\author{J.~I.~McDonald}
\email[]{jamie.mcdonald@uclouvain.be}
\affiliation{Centre for Cosmology, Particle Physics and Phenomenology,
Université catholique de Louvain,
Chemin du cyclotron 2,
Louvain-la-Neuve B-1348, Belgium}

\author{S.~Srinivasan }
\email[]{sankarshana.srinivasan@postgrad.manchester.ac.uk}
\affiliation{%
Jodrell Bank Centre for Astrophysics, School of Natural Sciences, Department of Physics and Astronomy, University of Manchester, Manchester, M13 9PL, U.K.
}

\label{firstpage}

\date{\today}

\begin{abstract}
 Axions are well-motivated candidates for dark matter. Recently, much interest has focused on the detection of photons produced by the resonant conversion of axion dark matter in neutron star magnetospheres. Various groups have begun to obtain radio data to search for the signal, however, more work is needed to obtain a robust theory prediction for the corresponding radio lines. In this work we derive detailed properties for the signal, obtaining both the line shape and time-dependence. The principal physical effects are from refraction in the plasma as well as from gravitation which together lead to substantial lensing which varies over the pulse period.  The time-dependence from the co-rotation of the plasma with the pulsar distorts the frequencies leading to a Doppler broadened signal whose width varies in time.  For our predictions, we trace curvilinear rays to the line of sight using the full set of equations from Hamiltonian optics for a dispersive medium in curved spacetime.  Thus, for the first time, we describe the detailed shape of the line signal as well as its time dependence, which is more pronounced compared to earlier results. Our prediction of the features of the signal will be essential for this kind of dark matter search.
\end{abstract}



\maketitle

\section{Introduction}

Understanding the nature and origin of dark matter remains one of the greatest challenges in contemporary particle physics and cosmology. One of the most compelling explanations is that dark matter is of a particle nature consisting of some as yet undiscovered cold (non-relativistic) component in the present-day Universe. Given that to date, the simplest WIMP candidates for dark matter have not yet been observed, attention is beginning to focus on other dark matter scenarios. A particularly popular explanation is that dark matter may be composed of light (pseudo) scalar fields corresponding to axions-like particles (ALPs) \cite{ref:misalign1,ref:misalign2,ref:misalign3}. These are  attractive both for their simplicity and ubiquity in beyond Standard Model theories \cite{Arvanitaki:2009fg, Svrcek:2006yi}.

The axion was initially introduced to explain the absence of charge-parity (CP) violation\footnote{Note that recently some authors have questioned whether the theory predicts CP-odd observables in the first place \cite{Ai:2020mzh}.} in Quantum Chromodynamics (QCD) \cite{ref:PQ, ref:K, ref:SVZ, ref:DFSZ, ref:Zhit} see also \cite{DiLuzio:2020wdo}. Regardless of its theoretical origin, one can still consider the possibility of dark matter consisting of generic ALPs with masses in the $\mu$eV range, interacting with photons via $\mathcal{L}_{a \gamma \gamma} = g_{a \gamma \gamma} a F_{\mu \nu} \tilde{F}^{\mu \nu}$, where $a$ is the axion field, and $F_{\mu \nu}$ and $\tilde{F}_{\mu \nu}$ are the photon field strength and its dual. 

One way to search for axion dark matter is by observing its decay into two photons \cite{ref:Sigl,Caputo:2018vmy,Caputo:2018ljp,Battye_2020,Carenza:2019vzg,Balkin:2020dsr,Buckley:2020fmh,Bernal:2020lkd,Caputo:2020msf,Fortin:2021cog,Nurmi:2021xds,An:2020jmf}. However, compact objects have long been known to offer a useful avenue in which to probe axions and ALPs in a variety of ways \cite{Raffelt:1996wa,Day:2019bbh,OHare:2020wum,Garbrecht:2018akc,Prabhu:2020yif,Edwards:2020afl,Fortin:2021sst,Poddar:2020qft,Harris:2020qim}. Neutron stars (NSs) in particular offer an exciting opportunity for increasing the possibility to detect axion dark matter by allowing axions to resonantly convert into radio photons in their magnetospheres. 

Historically, these ideas can be dated to a proposal by Pshirkov and Popov \cite{Pshirkov:2007st}, while the more general question of mixing of axions with photons in neutron star magnetospheres was also considered in \cite{ref:LaiHeyl}. More recently, the subject has seen a renewed interest \cite{ref:NS-Hook,ref:NS-Japan, Camargo:2019wou,  Safdi:2018oeu, Edwards:2019tzf, Battye_2020, Leroy:2019ghm}. 

One particularly useful aspect of radio detection of axion dark matter in NSs is that it has the potential to guide and complement  existing haloscope experiments. Notable proposed and ongoing examples are MADMAX \citep{Majorovits2017,Brun2019}, HAYSTAC \cite{Droster2019}, ADMX \cite{ref:ADMXf}, ORGAN \cite{McAllister2017} as well as many other proposed experiments involving novel condensed matter and metamaterial structures  \cite{Lawson:2019brd,Schutte-Engel:2021bqm,Baryakhtar:2018doz}.

At present, a number of studies have begun to take data to constrain the pulsar signal for axion dark matter  \cite{Foster:2020pgt,Darling:2020uyo,Darling:2020plz}. However, more work is required to properly characterise the shape and time-dependence of the radio line signal, including its (Doppler broadened) width, as investigated by us\footnote{We are indebted to G. Raffelt for originally bringing this possibility to our attention.} in \cite{Battye_2020}. The authors of \cite{Leroy:2019ghm} developed a ray tracing procedure for deriving more accurate observational properties of the signal. In the present work, we extend this analysis to self-consistently account for an inhomogeneous and time-dependent magnetosphere as well as including the effects of gravity. This allows us to incorporate the bending of the rays due to varying refractive index and compute the Doppler broadening of the signal from the time-dependence of the magnetosphere.


The remainder of this paper is organised as follows. In sec.~\ref{sec:axionconversion} and \ref{sec:ray_tracing} we review axion conversion in neutron star magnetospheres and describe how to trace rays through the magnetosphere in a space and time dependent plasma. In sec.~\ref{sec:signal} we present the signal properties resulting from our ray tracing analysis, including the shape of the Doppler broadened line signal and the effects of plasma in influencing the time-dependence of the signal. We finish in sec.~\ref{sec:discussion} where we offer our conclusions and suggestions for future work.

\section{Axion Conversion In neutron star magnetospheres}\label{sec:axionconversion}

We begin by introducing axion electrodynamics equations relevant for axion-photon mixing \cite{Battye_2020,Witte:2021arp,ref:NS-Hook}
\begin{align}
\square \, a + m_{\rm a}^2 a & = g_{\rm a \gamma \gamma}\textbf{E} \cdot \textbf{B}_0\,,\label{axionEOM}\\
\square \, \textbf{E} + \nabla (\nabla \cdot \textbf{E}) + \boldsymbol{\varepsilon} \cdot \ddot{\textbf{E}} &= -g_{\rm a \gamma \gamma} \ddot{a}\textbf{B}_0\,, 
\label{EOMPerts}
\end{align}
where $\textbf{B}_0$ is an external magnetic field, $\textbf{E}$ is the electric field of the photon and $a$ is the axion and $\varepsilon$ is the permittivity. In general the solutions to these equations are complicated by the geometry of the magnetic field and plasma in relation to the propagation direction of the incoming axion, as discussed in the 2D simulations \cite{Battye_2020} and the de-phasing arguments of \cite{Witte:2021arp}. This suggests a more comprehensive treatment of mixing is warrented in future work. However, if one specialises to the ``Planar case" case in which plasma gradients are aligned with the incident axion \cite{ref:NS-Hook}, and if one assumes the magnetic field is constant, the integration parameter $z$, then these equations simplify to 
\cite{Raffelt:1987im, ref:NS-Hook, Witte:2021arp}
\begin{eqnarray}
- \partial_z^2 \left(
\begin{array}{c}
E \\
a
\end{array}
\right)
= 
\left(
\begin{array}{cc}
\frac{\omega^2 - \omega_{\rm p}^2}{1 - \frac{\omega_{\rm p}^2}{\omega^2}\sin^2 \tilde{\theta}} & \frac{\omega^2 g_{a \gamma \gamma} B_0 \sin \tilde{\theta}}{1 - \frac{\omega_{\rm p}^2}{\omega^2}\sin^2 \tilde{\theta}} \\
\frac{\omega^2 g_{a \gamma \gamma} B_0 \sin \tilde{\theta}}{1 - \frac{\omega_{\rm p}^2}{\omega^2}\sin^2 \tilde{\theta}} & \omega^2 - m_a^2
\end{array}
\right)
\left(
\begin{array}{c}
E \\
a
\end{array}
\right)\,,
\end{eqnarray}
where $z$ is the arclength along the photon worldline, and $\tilde{\theta}$ is the angle between the magnetic field and the propagation direction, $B_0$ as the magnitude of the background magnetic field and $E$ is the electric field associated to the propagating photon and $\omega$ and $\omega_{\rm p}$ are the photon frequency and plasma mass, respectively. Resonant conversion occurs when $\omega_{\rm p} = m_a$ at some point $z = z_c$. This defines a  critical surface around the star on which $\omega_{\rm p} = m_a$. The conversion probability was derived in \cite{ref:NS-Hook} and reviewed in excellent detail\footnote{Note that following the release of this manuscript, a very recent and more extensive re-examination of axion-photon conversion in strongly magnetised plasmas has been given in \cite{Millar:2021gzs}.} in \cite{Witte:2021arp} using a method of stationary phase to arrive at
\begin{eqnarray}
 P_{\rm a \rightarrow \gamma}  = \frac{\pi}{2 (v^{a}_{\rm em})^2}(g_{a \gamma \gamma} B_\perp)^2 |\partial_z k_\gamma|^{-1}
\end{eqnarray}
where $B_\perp = B_0\sin \tilde{\theta}$ is the component of the magnetic field perpendicular to propagation and $k_\gamma = \sqrt{\omega^2 - \omega_{\rm p}^2}$ is the photon momentum and $v_{\rm em}$ is the axion velocity at emission\footnote{Formally, in a strongly magnetised medium the dispersion relation is $k_\gamma  = \sqrt{
\frac{\omega^2 - \omega_{\rm p}^2} {1 - \frac{\omega_{\rm p}^2}{\omega^2} \cos ^2 \tilde \theta  }}$ as pointed out in \cite{Witte:2021arp} (in contrast the dispersion used by some of the same author's in their earlier work \cite{Leroy:2019ghm} which used the isotropic dispersion relation above). This can introduce additional angular dependence as in \cite{Witte:2021arp}. These corrections can be incorporated into our pipeline in future work. However since the main goal of the present work is to understand the propagation of photons \textit{subsequent} to conversion so as to allow a like-with-like comparison to the straight line rays of ref.~\cite{Leroy:2019ghm} (re-produced in our fig.~\ref{fig:PulseProfile}), we shall use their expression  \eqref{eq:Probability}. }. This leads to a conversion probability
\begin{equation}\label{eq:Probability}
    P_{\rm a\rightarrow\gamma}  = \frac{\pi g_{a \gamma \gamma}^2 B_{\rm \perp}^2}{2 \,  \omega_{\rm p}'(\textbf{x}_{\rm em}) v^{a}_{\rm em}} \, ,
\end{equation}
also quoted in ref \cite{Leroy:2019ghm}, where $\omega_{\rm p}'(\textbf{x}_{\rm em}) = \hat{\textbf{k}}_{\rm em} \cdot \nabla \omega_{\rm p}$ is the projected plasma gradient onto the direction of propagation given by the unit vector $\hat{\textbf{k}}_{\rm em}$. Note this incorporates the full angular $\tilde{\theta}$ dependence of the conversion probability, which is contained implicitly in $B_\perp$ and the directional derivative, which are computed implicitly in our code.

At this point we make a comment about the form of the conversion probability, which results from a stationary phase approximation of the following integral
\begin{equation}\label{eq:Pintegral}
P_{\rm a \rightarrow \gamma} = \left| \int_{-\infty }^\infty dz^{\prime} \Delta_B (z^{\prime}) e^{\imath \int_0^{z^{\prime}} dz^{\prime\prime} \left[\Delta_\gamma(z^{\prime\prime}) - \Delta_{\rm a}(z^{\prime\prime})\right]} \right|^2\, ,
\end{equation}
where,
\begin{equation}
\Delta_{\rm a} = m_{\rm a}^2/2\bar{k}\,, \quad 
\Delta_\gamma = \omega_{\rm pl}^2/2\bar{k}\,, \quad 
\Delta_B = \omega g_{\rm a\gamma \gamma} B_0/2\bar{k}\,.
\end{equation}
 Note that $\bar{k}$ is the local average momentum associated to the average of the eigevalues of the mixing matrix (see appendix B in reference \cite{Battye_2020} for a more detailed explanation).

 The method of stationary phase is essentially a Gaussian approximation of the exponent, where the width of the Gaussian is set by the second derivative of the phase $\Delta_{\rm p}'$ giving the $\omega'_{\rm p}$ in the denominator of $P_{a \rightarrow \gamma}$. However, when rays are tangent to the critical surface, which is the case at the edge of the image, the projected derivative $\hat{\textbf{k}}_{\rm em} \cdot \nabla \omega_{\rm p}$ vanishes, so that $1/\omega_{\rm p}'$ becomes singular. It might be tempting to interpret this as the limiting case of strongly adiabatic evolution, as discussed in \cite{Battye_2020}. However, such points actually represent a breakdown of the simplest treatment of stationary phase. Points where the first \textit{and} second derivative of the phase vanish are known as \textit{degenerate stationary points} and are the subject of study in their own right. The perturbative computation of the conversion probability results in the following integral

 Computing the integral in \eqref{eq:Pintegral} for degenerate stationary points is a subtle topic intimately connected to the field of catastrophe theory \cite{poston1996catastrophe}. We remark that if one expands the exponent to higher order, one obtains an exponent of the form $\sim z^2 \Delta'_{\gamma} + z^3 \Delta''_\gamma$. Integrals with cubic exponents of this form are expressible in terms of Airy functions, which are again ubiquitous in optics applications of catastrophe theory. The stationary phase approximation for non-degenerate ($\Delta'_\gamma \neq 0$) points, is valid so long as the first term is dominant over the second, this can be characterised by the dimensionless quantity $(\Delta'_\gamma )^2/(\Delta''_\gamma)^{4/3}$. This is in fact the only dimensionless quantity one can construct from $\Delta_{\gamma}'$ and $\Delta_{\gamma}''$. In the present work, we take a precautionary approach, and excise any points from our analysis for which have $(\Delta'_\gamma )^2/(\Delta''_\gamma)^{4/3} < 1$. These typically occur near the edge of the image where rays become tangent to the critical surface. We leave a more careful treatment of degenerate stationary points for future work.

Returning to our discussion, we can now use eq.~\eqref{eq:Probability} to compute the radiant intensity $I_{\rm em}$ at the point of emission from the critical surface:
\begin{equation}\label{eq:Iem}
I_{\rm em}(\textbf{x}_{\rm em}, \textbf{v}_{\rm em}^a)= \frac{1}{4 \pi} m_a v^a_{\rm em} \rho_{\rm DM}(\textbf{x}_{\rm em}) P_{a \rightarrow \gamma }(\textbf{x}_{\rm em}, \textbf{v}^a_{\rm em})\,,
\end{equation}
where $1/4 \pi$ is the fraction of axions emitted in a particular direction for an isotropic axion distribution at the critical surface. This gives the energy flow per solid angle for a given point on the critical surface. Note the dark matter density is enhanced at the critical surface by a factor \cite{Alenazi:2006wu,ref:NS-Hook,Leroy:2019ghm} 
\begin{equation}\label{eq:Liouville_Grav_Increase}
    \rho_{\rm DM}(\textbf{r})  =\frac{2}{\sqrt{\pi} }\frac{1}{v_0} \sqrt{\frac{2GM}{r}} \rho^{\infty}_{\rm DM}\,,
\end{equation}
where $\rho^\infty_{\rm DM}$ is the dark matter mass-density at infinity, $M$ is the mass of the neutron star, and $v_0$ is the velocity dispersion of dark matter.

\begin{figure}
     \centering
     \includegraphics{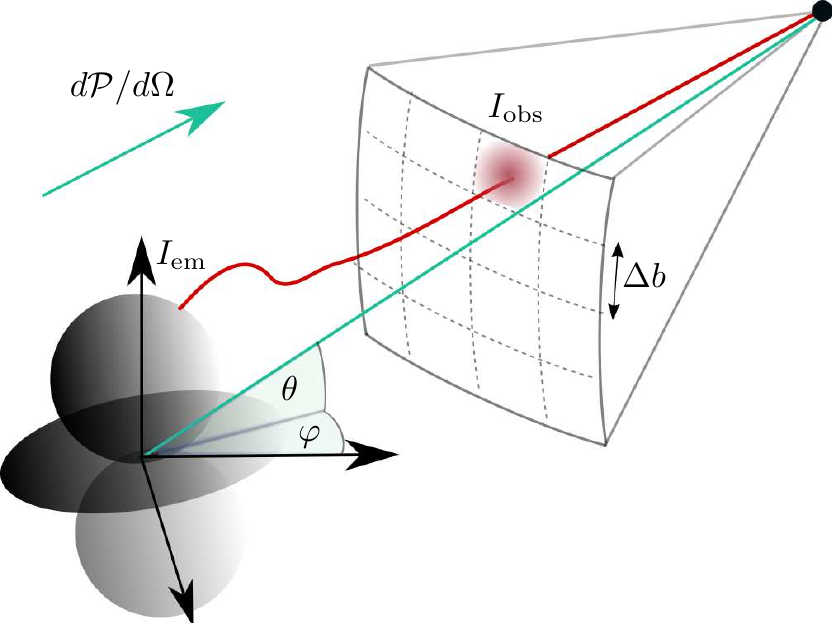}
     \caption{\textbf{Ray tracing geometry.} A sketch of the ray tracing scheme. The image plane is divided into pixels side length $\Delta b$ and lies perpendicular to the line of sight  $(\theta, \varphi)$ in polar coordinates defined about the centre of the star. A ray (red) is back-traced from the center of each pixel towards the star. If the ray hits the critical surface, it is assigned an appropriate intensity $I_{\rm em}$ according to the probability for axion-photon conversion. }
     \label{fig:ray_Cartoon}
 \end{figure}

Assume we observe the neutron star from a particular line of sight $(\theta,\varphi)$, defined relative to the center of the star, with the north pole aligned with the rotation axis. Consider the set of angles $(\theta_{\rm obs},\varphi_{\rm obs})$  covering an observation region of angular size $\Delta \Omega$ containing the image of the magnetosphere. The total collected flux $F_{\rm obs}$ from this region is given by
 \begin{equation}\label{eq:flux}
   F_{\rm obs}(\theta,\varphi) =\int_{\Delta \Omega}  d\Omega(\theta_{\rm obs},\varphi_{\rm obs}) \,  I_{\rm obs} (\theta_{\rm obs}, \varphi_{\rm obs})\,.
 \end{equation}
The area $dA$ subtended by $\Delta \Omega$ is 
 \begin{equation}
     d A = D^2 \Delta \Omega , 
 \end{equation}
 where $D$ is the distance between the observer and the star. We can divide this area into pixels characteristic size $\Delta b$, labelled by $i$, with area $dA_i = (\Delta b)^2$ and infinitesimal solid angle $\Delta \Omega_i$ so that $dA_i = (\Delta b)^2 =  D^2 \Delta \Omega _i$. The integral \eqref{eq:flux} can then be approximated by 
 \begin{eqnarray}
 F_{\rm obs}(\theta, \varphi) = \frac{1}{D^2} \sum_{i} (\Delta b)^2 I_{\rm obs}^i(\theta^i_{\rm obs}, \varphi^i_{\rm obs}) \,.
 \end{eqnarray}
 This scenario is sketched in Fig.~\ref{fig:ray_Cartoon}. 
 Thus, to calculate the flux, we backtrace rays onto the critical surface. The specific intensity at the point of emission can then be related to that at detection  (see appendix \ref{appendix:RadiativeTransport} or \cite{Rogers:2015dla} and references therein) by using the property
  \begin{equation}
     \frac{I}{n^2 \omega^3} = \text{constant} \,,
 \end{equation}
 along rays where $n$ is the refractive index.  Hence we have the following relation
 \begin{eqnarray}\label{eq:IemeqlIobs}
    \frac{I_{\rm obs}}{n^2_{\rm obs} \omega^3_{\rm obs}} = \frac{I_{\rm em}}{n^2_{\rm em} \omega^3_{\rm em}} \,,
\end{eqnarray}
where $n_{\rm em}$ and $n_{\rm obs}$ and $\omega_{\rm em}$ and $\omega_{\rm obs}$ etc.~are measured in a coordinate system at rest with respect to the star. 
Combining this relation together with the definition \eqref{eq:Iem} for the intensity at emission and using the fact that far from the star $n_{\rm obs} \simeq 1$  we arrive at 

\begin{equation}\label{eq:Ffinal}
        F = \frac{1}{D^2} \sum_{i}  \frac{(\Delta b)^2}{(n^i_{\rm em})^2 } \tilde{f}(r_{\rm em}, r_{\rm s}) \frac{\rho_{\rm DM}(\textbf{x}^i_{\rm em}) v_{\rm em}^a P_{\rm a \rightarrow \gamma}}{4\pi}\, , 
 \end{equation}
 where $\tilde{f}(r, r_{\rm s}) = \omega_{\rm obs}^3/\omega_{\rm em}^3 = (1 - r_s/r_{\rm em})^{3/2}\lesssim 1 $ is the gravitational red-shift factor.  Hence in comparison to the formula in \cite{Leroy:2019ghm}, we have an additional red-shift factor of order 1, and a factor of $1/n_{\rm em}^2$ which compensates for the shrinking of the image in the image plane due to plasma lensing discussed in the next section. Finally we can relate the radiated power $d\mathcal{P}/d\Omega$ to the flux $F_{\rm obs}$ via
 \begin{equation}
 F_{\rm obs} = \frac{1}{D^2}\frac{d \mathcal{P}}{d \Omega}\,.
 \end{equation}

\section{Ray Tracing}\label{sec:ray_tracing}

Ray tracing is a powerful technique for understanding the emission properties of astrophysical bodies and enables one to track the position, frequency and momentum of photons. The rays then contain all the information required to reconstruct the image of the object, as well as giving the angular power dependence, lensing effects and frequency distortions. Such techniques have been applied to both stars \cite{Vincent:2017emv} and black holes \cite{Psaltis:2010ww,Johannsen:2010ru,Rogers:2015dla}, where in the latter case,  the rays are geodesics of the spacetime metric rather than the plasma, but the principle is the same. 

Ray tracing techniques were first applied to axion dark matter conversion in \cite{Leroy:2019ghm}. Here, the authors back-traced straight-line rays from the observer to the critical surface on which photons are produced, matching each ray onto its corresponding conversion amplitude. In this work, we examine how plasma affects ray tracing and the corresponding properties of the radio lines. 

We shall see that the plasma has two very important effects which greatly modify signal properties in two important ways. The first is a new time-dependence resulting from the refraction of rays, which causes stronger pulsing of the signal. The plasma acts as a time-dependent lens, causing a variation in the number of rays reaching the observer, which fluctuates over the pulse period. 
The second effect involves the variation of frequency along rays, which allows us to determine the Doppler broadening of each photon, and by summing over all rays, derive the exact line shape of the signal in frequency space.


Our starting point for understanding the propagation is similar to the discussion found in \cite{Rogers:2015dla}, which is a generalisation of the flat space case discussed in \cite{Weinberg1962} (and applied in refs.~\cite{McDonald:2020why,McDonald:2019wou}) which describes the propagation of photons in an inhomogeneous and time-dependent plasma using Hamiltonian optics. 
We begin with a dispersion relation in a cold, isotropic plasma


\begin{equation}
    g_{\mu \nu} k^\mu k^\nu +  \omega_{\rm p}^2 =0\,,
\end{equation}
where $g_{\mu \nu}$ is the space time metric. Taking covariant derivatives of this equation, we arrive at
\begin{eqnarray}\label{eq:geodesic0}
k^\nu \nabla_\nu k^\mu = - \frac{1}{2}\partial^\mu \omega_{\rm p}^2, 
\end{eqnarray}
where we used $\nabla_\mu k_\nu = \nabla_\nu k_\mu$ since $k_\mu  = \partial_\mu \Theta$ is the derivative on the eikonal phase $\Theta$ in the relevant WKB approximation. We then define worldlines $x^\mu(\lambda)$ associated to these rays satisfying
\begin{equation}\label{eq:geodsic1}
\frac{d x^\mu}{d \lambda} = k^\mu,
\end{equation}
where $\lambda$ is an arbitrary worldline parameter. Putting this together we arrive at
\begin{eqnarray}\label{eq:geodesic2}
\frac{d^2 x^\mu}{d \lambda^2} + \Gamma^\mu_{\nu \rho} \frac{dx^\nu}{d\lambda} \frac{dx^\rho}{d\lambda} = - \frac{1}{2} \partial^\mu \omega_{\rm p}^2\,, 
\end{eqnarray}
where $\Gamma^\mu_{\nu \rho}$ are the Cristoffel symbols associated to the connection. We interpret the spatial components of the plasma derivatives on the right hand side of eq.~\eqref{eq:geodesic2} as an effective force, leading to the refraction of rays. Meanwhile, the temporal derivatives lead to frequency evolution along the worldline, as is apparent from eq.~\eqref{eq:geodesic0}.

We choose a simple Schwarzschild metric to model the star's gravitational field
\begin{eqnarray}
ds^2  
= - f(r) dt^2 +\frac{dr^2}{f(r)} + d\Omega^2\,,
\end{eqnarray}
where $f(r) = 1- r_s/r$ and $r_s = 2GM$ is the Schwarzschild radius for a neutron star of mass $M$. The refractive index $n$ of the medium is defined by
\begin{equation}\label{eq:refractiveIndex1}
    n^2 = 1 - \frac{\omega_{\rm p}^2}{\tilde{\omega}^2}\,,
\end{equation}
where $\tilde{\omega}$ is the frequency in a coordinate system at rest with respect to the neutron star and includes the gravitational red-shift. 
\begin{equation}\label{eq:redShiftFreq}
\tilde{\omega}(r) = \left(1 - \frac{r_s}{r}\right)^{- 1/2} \omega\,.
\end{equation}
Here $\omega=-k_0$ is the co-moving frequency related to the temporal component of $k_\mu$. In the absence of time-dependence in the plasma, $\omega$ is conserved along rays. 
\subsection{Doppler Broadening}\label{sec:Doppler}
When the plasma is time-dependent, $\omega$ evolves along rays according to eq.~\eqref{eq:geodesic0}, which can be re-written in terms of coordinate time as
\begin{eqnarray}\label{eq:DopplerEvolution}
\frac{d \omega}{d t} = - \frac{f}{2 \omega}\partial_t \omega_{\rm p}^2\,.
\end{eqnarray}
 The key point for this work, is to note that when the plasma background is time-dependent, as happens for the plasma around a neutron star, the photon frequency evolves according to eq.~\eqref{eq:DopplerEvolution}. We can obtain an estimate for the Doppler broadening which gives a frequency shift $\delta \omega$ satisfying
\begin{eqnarray}\label{eq:DopplerBroadening1}
\delta \omega \simeq \frac{1}{2 \omega } \,\int  dt \, \partial_t \omega_{\rm p}^2(t,\textbf{x}_0(t))\,, 
\end{eqnarray}
where $\omega$ is frequency before broadening, and $\textbf{x}_0$ gives the ray worldline. Formally we solve eq.~\eqref{eq:DopplerBroadening1} numerically for each ray, allowing us to build up the exact line shape for a given magnetosphere model.


\subsection{Effect of Gravity}\label{sec:grav}
When taken in combination with refractive plasma effects, gravity plays an important role in influencing the characteristic size of the lensed image of the star. To understand how this happens, it is instructive to consider a simple spherically symmetric and stationary plasma with $\omega_{\rm p} = \omega_{\rm p}(r)$. In this case, one has two conserved quantities: the frequency $\omega_{\rm obs}$ and angular momentum $L$ which is related to the impact parameter $b$ by $L = \omega b$. One can derive a simple energy conservation equation for the radial coordinates \cite{Rogers:2015dla} corresponding to motion in an effective potential
\begin{eqnarray}
\left(\frac{d r}{d \lambda}\right)^2 = \omega^2_{\rm obs} -  \Bigg(1 - \frac{r_s}{r}\Bigg)\Bigg( \frac{\omega^2_{\rm obs} b^2}{r^2} + \omega_{\rm p}^2(r)\Bigg)\,.
\end{eqnarray}
Back-traced photons which are capable of reaching the critical surface, must have a distance of closest approach $r_{\rm min}$ satisfying $r_{\rm min} \leq r_c$, where $r_c$ is the radius of the critical surface. Since $r_{\rm min}$ is by definition a stationary point along the geodesic at $r=r_{\rm min}$ we must have $dr/d\lambda =0$. The maximum impact parameter $b_{\rm max}$, corresponds to those rays which just skim the critical surface. For these rays $r_{\rm min} = r_c$. We therefore have the maximum impact parameter for rays which can reach the critical surface
\begin{equation}\label{eq:bmax}
b_{\rm max} = r_c \left[\frac{1} {1 - r_s/r_c} -  \frac{m_a^2}{\omega^2_{\rm obs}} \right]^{1/2}\,,
\end{equation}
where we used the definition that at the critical surface, $\omega_{\rm p}^2(r_c) = m_a^2$. The key point to note is that $b_{\rm max}$ sets the characteristic size of the image in the image plane, which in the toy example we describe here, is a circle radius $b_{\rm max}$.

We also know that since $\omega_{\rm obs}$ is the asymptotic frequency of photons, it satisfies $\omega_{\rm obs }^2 \simeq m_a^2(1 + v_0^2) $ where $v_0$ is the asymptotic velocity of the axion, set by the velocity dispersion of dark matter.  Putting this together, we see that to leading order in $v_0$ and $r_s$, we have that the characteristic size of the image is given by $r_{\rm image } \simeq b_{\rm max}$
\begin{equation}\label{eq:imagesize}
r_{\rm image} \simeq r_c \left[ \frac{r_s}{r_c} + v_0^2\right ]^{1/2}\,.
\end{equation}
Let us now consider the scales at play. We can use a canonical model for the plasma density \cite{ref:GJ} (see eq.~\eqref{eq:rcExp})  to estimate the size of $r_c$ by equating $\omega_{\rm p}(r_c) = m_a$. For the pulsar J0806.4-4123 used in this work and \cite{Leroy:2019ghm}, we have $B_0 = 2.5 \times 10^{13}\, {\rm G}$ and $P = 11.37\, {\rm sec}$ so that for a mass $m_a = 0.5\, \mu \text{eV}$ we obtain a characters tic radius $r_c \simeq 5 R$. For a neutron star of mass $M= M_{\odot}$ and radius $R=10{\rm km}$ we have $r_s/R \simeq 0.3$. Note this value is quite high, owing to neutron stars being very compact, and quite close to being black holes. 

\begin{figure*}
    \centering
    \includegraphics[scale=0.7]{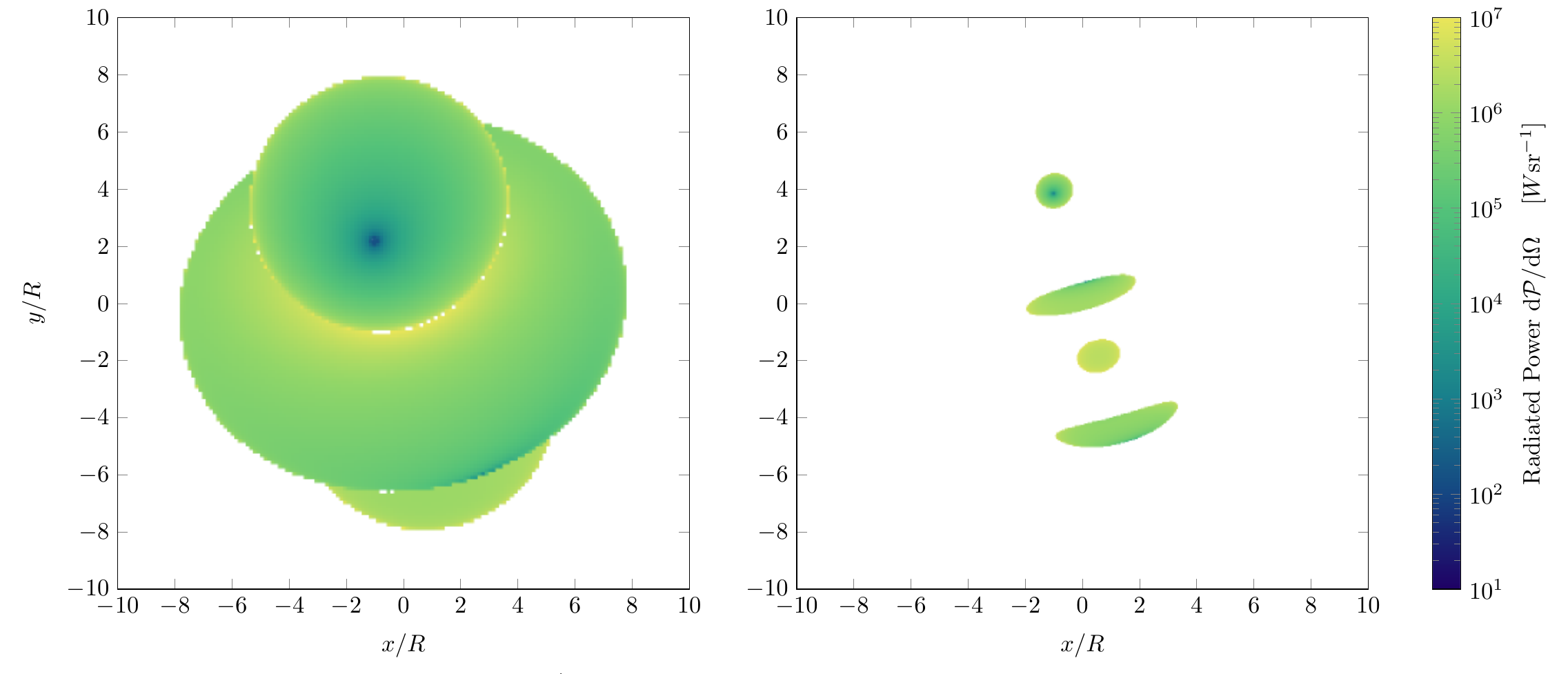}
    \caption{\textbf{Radiated power in image plane: vacuum vs.~plasma.} The view of radio photons at frequency $\omega \simeq m_a$ produced by axion dark matter conversion, as seen in the image plane perpendicular to an observing angle $\theta = 36^\circ$.  We chose model parameters compatible with observations of  NS J0806.4-4123, with $B_0 = 2.5 \times 10^{13}$G, period $P = 11.37 {\rm sec}$ and magnetic misalignment angle $\alpha = 18 ^\circ$. The axion mass chosen was $m_a = 0.5 \mu$eV with a coupling $g_{a \gamma \gamma} = 10^{-12} $GeV$^{-1}$. The left image results from straight-line rays propagating through vacuum, reproducing the results of ~\cite{Leroy:2019ghm}. The right panel shows the presence of novel plasma effects considered in this work, computed via eq.~\eqref{eq:geodesic2}. The time evolution of the  image in the plasma case can be viewed  \href{https://www.dropbox.com/s/3fizit62zy3mfz8/Image_Plasma.mp4?dl=0}{here} and can be contrasted against the \href{https://youtu.be/VyA1-qbIqB4}{vacuum case} derived in \cite{Leroy:2019ghm}.}
    \label{fig:ImagePlaneComparison}
\end{figure*}

Coming back to eq.~\eqref{eq:imagesize}, we see that for the NS values chosen above, $r_s/r_c\simeq 0.06 \gg  v_0^2 \sim 10^{-6}$ so that gravity plays a vital role from a ray tracing perspective, in that it makes the area of the image $\mathcal{A}_{\rm image} \sim r_{\rm image}^2 $ four orders of magnitude larger in comparison to a plasma analysis in flat space! Numerically, speaking, this greatly simplifies the task of locating and resolving the image of the critical surface in the image plane. Locating a larger region, is of course much easier than hunting for a highly lensed ``pin prick". There is of course no such issue for the straight-line rays considered in \cite{Leroy:2019ghm} where the characteristic size of the image was just given by the geometric cross-section $\mathcal{A}^{\rm vac}_{\rm image} \sim r_c^2$. It is interesting to note that the authors of \cite{Leroy:2019ghm} claimed that gravity in the absence of plasma only produces a small percent-level correction to the total power. However, as explained above, when taken in combination with plasma, gravity in fact becomes an important component in making the problem numerically tractable by counter-balancing strong refraction from the magnetosphere. The relative image sizes with and without plasma can be seen by comparing the two panels in Fig.~\ref{fig:ImagePlaneComparison}.

The interpretation of this tension between gravity and plasma is actually rather straightforward. Plasma is repulsive, and so tends to deflect rays away from the observer, which would otherwise reach the image plane in the vacuum case. Gravity meanwhile is attractive, counteracting the effects of plasma, leading to a larger image in the image plane. These distinctions are of course vitally important when it comes to accurate ray tracing, which is crucial both to derive the line shape and time-dependence of the signal. 

At this point, it is also interesting to relate the expression \eqref{eq:bmax} to the refractive index appearing in eq.~\eqref{eq:Ffinal} which can be read off from eq.~\eqref{eq:refractiveIndex1} and eq.~\eqref{eq:redShiftFreq} as
 \begin{equation}
    n^2_{\rm em} =\frac{1}{1-r_s/r_c}\left(1- \frac{r_s}{r_c} - \frac{m_a^2}{\omega^2_{\rm obs.}} \right)\,,
\end{equation}
where we have used the fact that $\omega_{\rm p}^2 = m_a^2$ at the critical surface. 
using $\omega_{\rm obs}^2 \simeq m_a^2 (1+ v_0^2)$ we can write
\begin{equation}
    n^2_{\rm em} \simeq \frac{1}{1-r_s/r_c}\left(\frac{r_s}{r_c} + v_0^2 \right)\,.
\end{equation}
Note therefore that $n_{\rm em} \ll 1$. It is interesting to relate this result to the effective size of the image as given by eqs.~\eqref{eq:bmax} and \eqref{eq:imagesize}, from which we can now read off
\begin{equation}
 \mathcal{A}_{\rm image} \propto r_c^2 n_{\rm em}^2 \,. 
\end{equation}
 Hence although the image size is shrunk relative to the vacuum by a factor $n_{\rm em}^2 \ll 1$, the rays are more intense, receiving a compensating enhancement $1/n_{\rm em}^2$. This is just the lensing principle: that focused light is brighter, so that one ends up with a smaller, but brighter image relative to the vacuum case. As a result, the total power in all directions is conserved when plasma effects are switched on and off but crucially the angular power distribution and time-dependence are different. 
 
 Underlying this is of course a conservation principle: so long as attenuation in the medium is neglected, the total power integrated over all emission directions from the magnetosphere must be the same with and without plasma effects. This is because refraction only bends rays, reassigning power to different outgoing directions relative to the vacuum case.

\subsection{Ray Tracing Procedure}

 Our overall procedure adds many new features to the analysis performed in ~\cite{Leroy:2019ghm}. As emphasised elsewhere in the text, we include new refractive effects whereas ~\cite{Leroy:2019ghm} considered straight line rays in vacuum. We therefore must solve a differential equation for each ray which can be highly refracted in the magnetosphere affecting the time dependence of the signal.

Furthermore, we are able to quantitatively assess the Doppler broadening of rays for the first time. For each ray, we compute the Doppler broadening as follows.
For a given frequency $\omega \simeq m_a$ we first compute a ray back-traced from a given pixel. Rays from the same pixel with similar frequencies $|\omega - m_a|/m_a \lesssim v_0^2 \sim 10^{-6} $ have the same world lines up to negligible corrections $v_0^2 \sim 10^{-6}$. Hence, for a given pixel, all frequencies of interest can be treated as being transported along the same ray to excellent approximation. All that then remains is to compute the frequency shift along that ray. Yet again, the different frequencies in this range experience the same frequency shift $\delta \omega$ up to \textit{relative} corrections $\mathcal{O}(v_0^2)$.  This can be seen by examining the denominator of eq.~\eqref{eq:DopplerBroadening1}. Consider a spread of frequencies at the point of emission $\omega_{\rm em} = m_a + \delta \omega_{\rm em}$ where $\delta \omega_{\rm em} \lesssim v_0^2\, m_a$. The frequency dependence appears in the denominator of the evolution equation, which can be Taylor expanded as $1/\omega \simeq 1/m_a(1 + \delta \omega/m_a + \cdots)\simeq 1/m_a + \mathcal{O}(v_0^2)$ so that the frequency shift at the point of detection satisfies
\begin{align}
\delta \omega_{\rm obs} & \simeq \left[ \frac{1}{2m_a } + \mathcal{O}(v_0^2)\right]\,\int  dt \, \partial_t \omega_{\rm p}^2(t,\textbf{x}_0(t)) \nonumber \\
&\simeq \frac{1}{2 m_a}\int  dt \, \partial_t \omega_{\rm p}^2(t,\textbf{x}_0(t))
\end{align}
meaning that the Doppler broadening for a given ray is essentially achromatic within the relevant frequency range. Thus, the reason different rays experience different Doppler broadening is driven by their taking different paths $\textbf{x}_0(t)$ through the magnetosphere and ending on different points in the observing plane, rather than their different frequencies. Therefore, for a given  ray, we can essentially apply an identical Doppler shift to all frequencies. As a result, there is no need to undertake an intractable back-scanning over a large number of observing frequency bins. Such an approach would only correct what is already seen in Fig.~\ref{fig:LineShapes} to one part in a million.

We also use an adaptive method to locate the image of the critical surface, which is much smaller than the unlensed version in ~\cite{Leroy:2019ghm} where such a procedure would not be necessary since straight line rays simply produce the geometric cross-section of the critical surface. Our explicit algorithm is as follows.

\begin{enumerate}
    \item As shown in Fig.~\ref{fig:ray_Cartoon}. we begin by choosing a given observing direction $(\theta, \varphi)$. This gives the line of sight connecting the observer to the origin at the centre of the star. One then constructs an image plane perpendicular to this line of sight. This plane is then divided into pixels of side length $\Delta b$. A light ray emanates from the centre of each. This light ray is then back-propagated from the image plane using eq.~\eqref{eq:geodesic2}.  As rays are back-propagated one either records a ``hit" or a ``miss" dependent on whether the ray reaches or is deflected from the critical surface.  
    
    \item We then use the following adaptive method to resolve the image. First we perform a coarse-grained search with a larger pixel size to locate the disjoint regions in the image plane. Once found, these are then re-scanned with a smaller pixel size. This saves an inefficient high resolution scan over the entire image plane, most of which contains misses due to the image being lensed to a small, high intensity region, as can be seen in Fig.~\ref{fig:ImagePlaneComparison}.
    
    \item These rays can then be assigned the corresponding radiant intensity \eqref{eq:Iem} at the point of emission on the critical surface. The total luminosity then follows from integrating over the image plane, given by summing up the intensity carried by each ray and multiplying by $(\Delta b)^2$ according to eq.~\eqref{eq:Ffinal}.
    
    \item To compute the Doppler broadening, we then take these rays and evolve the frequency evolution along each ray according to eq.~\eqref{eq:DopplerEvolution}. This enables us to see how a line signal with frequency $\omega  \sim m_a$ is broadened due to each ray $i$  receiving a correction $\omega \rightarrow \omega  + \delta \omega $. The final frequencies at the point of detection can then be binned, so as to derive the exact shape of the signal in the frequency domain.

\end{enumerate}

    Our code is written in Mathematica using its in-built differential equation solvers and event locators to detect the critical surface. We also made use of stiff solver options which are required for especially oscillatory worldlines which are multiply reflected. Even though the solver uses an adaptive step size, one must also be sure to choose a maximum integration time-step along rays which is sufficiently small to achieve good convergence of results.  In the future we hope to make our code publicly available. 
   
\begin{figure*}[t]
    \centering
    \includegraphics[width=0.9\textwidth]{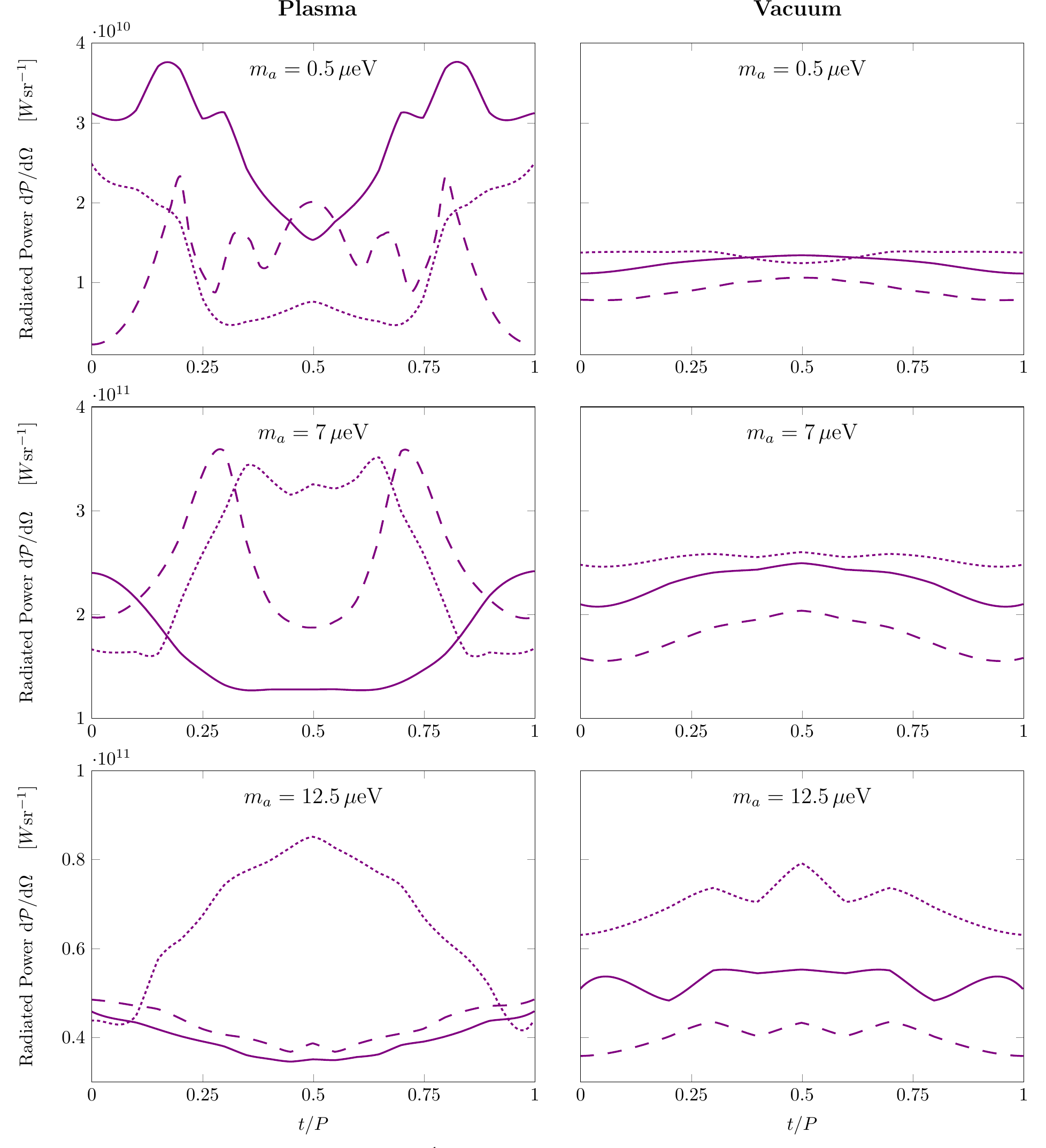}
    \caption{\textbf{Pulse profiles.} A comparison of pulse profiles with and without plasma is shown in the left and right columns, respectively. The plasma case is computed via eq.~\eqref{eq:geodesic2}. Plots show the radiated power over all frequencies. From top to bottom we show the results for axion masses $m_a = 0.5, \, 7, \, 12.5\, \mu \text{eV}$.  The dashed, solid and dotted lines correspond to an observing polar angle of $\theta = 36^\circ, 54 ^\circ $ and $72^\circ$, respectively. 
   Other model parameters are chosen as in Fig.~\ref{fig:ImagePlaneComparison}.
    }
    \label{fig:PulseProfile}
\end{figure*}

  We computed rays in parallel on 32-core cluster nodes. A full pulse profile corresponds to one of the curves in fig.~\ref{fig:PulseProfile}. The time taken to compute one of these depends on the following factors (i) the resolution required, set by the total number of pixels: typically tens of thousands (ii) the number of time-steps sampled over the pulse period:  more are required to resolve highly oscillatory pulse profiles (iii) the number of 32-core nodes available, which for our purposes was three (iv) the maximum integration time-step used along each ray. Another positive is that the performance of our procedure can be improved even further by implementing our algorithm in another language, e.g. Julia, C++ etc.~where the in-built solvers are significantly faster.



\section{Signal Properties}\label{sec:signal}

In order to make a phenomenological prediction for the signal, we take a simple Goldreich-Julian (GJ) model for the plasma density \cite{ref:GJ}. This begins with the magnetic field of an inclined rotating dipole \cite{Petri:2016tqe}
\begin{align}
B_r &= B_0 \left(\frac{R}{r}\right)^3 \left(\cos\chi\cos\theta + \sin\chi\sin\theta\cos\psi\right)\,, \nonumber  \\
B_\theta &= \frac{B_0}{2} \left(\frac{R}{r}\right)^3 \left(\cos\chi\sin\theta - \sin\chi\cos\theta\cos\psi\right)\,, \nonumber \\
B_\phi &= \frac{B_0}{2} \left(\frac{R}{r}\right)^3 \sin\chi \sin\psi \, 
\end{align}
where $\chi$ is the angle between the rotation axis and magnetic dipole moment and $\psi(t) = \phi - \Omega \,t$. The neutron star has a surface magnetic field $B_0$, radius $R$ and rotational frequency $\Omega = 2\pi/P$ where $P$ is the period. The GJ model gives the density of charge carriers
\begin{align}
n_{\mathrm{GJ}}(\mathbf{r}) = \frac{2\, \boldsymbol{\Omega} \cdot \mathbf{B}}{e} \frac{1}{1 - \Omega^2 \,r^2\, \sin^2 \theta}\,,
\label{eq:nGJ}
\end{align}
where $\boldsymbol{\Omega} = \Omega \hat{z}$ is the constant NS rotation vector. Neglecting the relativistic terms in the denominator, we arrive at
\begin{align}\label{eq:GJDensity}
n_{GJ} 
=\frac{B_0 \Omega}{2 \, e} \left( \frac{R}{r}\right)^3\left[ 
\cos \chi + 3 \cos \chi \cos(2 \theta) + 3 \sin \chi \cos \psi  \sin 2 \theta
\right]   .
\end{align}
 The plasma frequency is then given by
\begin{equation}
\omega_{\rm p} = \sqrt{ \frac{4 \pi \, \alpha_{\rm EM} \, \left| n_{\rm GJ} \right|  }{m_{\rm e}} },
\end{equation}
where $\alpha_{\rm EM} = e^2 /4 \pi$ is the fine structure constant and $m_{\rm e}$ is the electron mass\footnote{Note that here we assume two charge-separated regions consisting of positrons and electrons, as in \cite{ref:NS-Hook}. However, as pointed out in \cite{Safdi:2018oeu}, if the positively charged region consists of \textit{ions} the plasma frequency experiences the replacement $m_e \rightarrow m_p$ which can alter the location of the critical surface when equating $m_{\rm a} = \omega_{\rm p}$. }. We emphasise that our algorithm can be easily applied to any magnetosphere model in future work. 
We now go on to discuss two observationally relevant properties of the signal: Doppler broadening (which gives the frequency dependence of the signal) and the time-dependence of the signal amplitude as characterised by its pulse profile. Both these are relevant from an observational standpoint in that they play a role in determining sensitivity to the axion-photon coupling.

\subsection{Time dependence} 

We can characterise the time-dependence of the signal via the relative variance, defined by
\begin{eqnarray}
\sigma = \frac{\braket{(d\mathcal{P}/d\Omega  )^2 }}{ \braket{d\mathcal{P}/d\Omega} ^2} - 1, 
\end{eqnarray}
where $\braket{\cdots}$ denotes pulse averaging. The time-dependence of the signal for various parameter choices is displayed in table \ref{tab:timeDep}. Typically the time-dependence is stronger for lower axion masses, where it tends to also be larger than the variance reported in the vacuum case. This is also reflected in the comparison between the left and right columns in Fig.~\ref{fig:PulseProfile} which displays the pulse profile of the radiated power. We take as benchmark values the three axion masses and observing angles used in \cite{Leroy:2019ghm}. Therefore, we see that plasma enhances the time-dependence of the signal.

\begin{table}
\centering
\small%
\begin{tabular}{p{1.5cm}|p{1.5cm}|p{1.5cm}|p{1.5cm}}
$m_a$  & angle $\theta$  & $\sigma_{\rm vac.}$ ($\%$) & $\sigma_{\rm plas.}$ \, ($\%$) \\
\hline
\hline
0.5 $\mu$eV & $36^\circ$ & 1.2 & 20.0 \\

                 & $54^\circ$ & 0.4  & 5.7 \\
 
                 & $72^\circ$ & 0.1  & 34.0 \\
\hline
7 $\mu$eV & $36^\circ$ & 0.9 & 5.1 \\

                 & $54^\circ$ & 0.4 & 6.4 \\
 
                 & $72^\circ$ & 0.0 & 8.7 \\
\hline
12.5 $\mu$eV & $36^\circ$ & 0.4 & 0.8 \\

                 & $54^\circ$ &0.2 & 0.8 \\
 
                 & $72^\circ$ & 0.4 & 4.8 \\
\end{tabular}
\caption{\textbf{Signal time-dependence.} The relative time-variance of the signal $\sigma = \braket{(d\mathcal{P}/d\Omega)^2}/\braket{d\mathcal{P}/d\Omega}^2 - 1$ where $\braket{\cdots}$ denotes pulse averaging. The quantities $\sigma_{\rm vac.}$ and $\sigma_{\rm plas}$ respectively give the variance with and without plasma refraction. We displace results for different observing angles $\theta$ and axion masses $m_a$. The magnetic alignment angle was $\chi = 18^\circ$ with other model parameters as in previous plots.  }
\label{tab:timeDep}
\end{table}

\subsection{Doppler Broadening} 
\begin{figure}[t]
    \centering
      \includegraphics[scale=0.7]{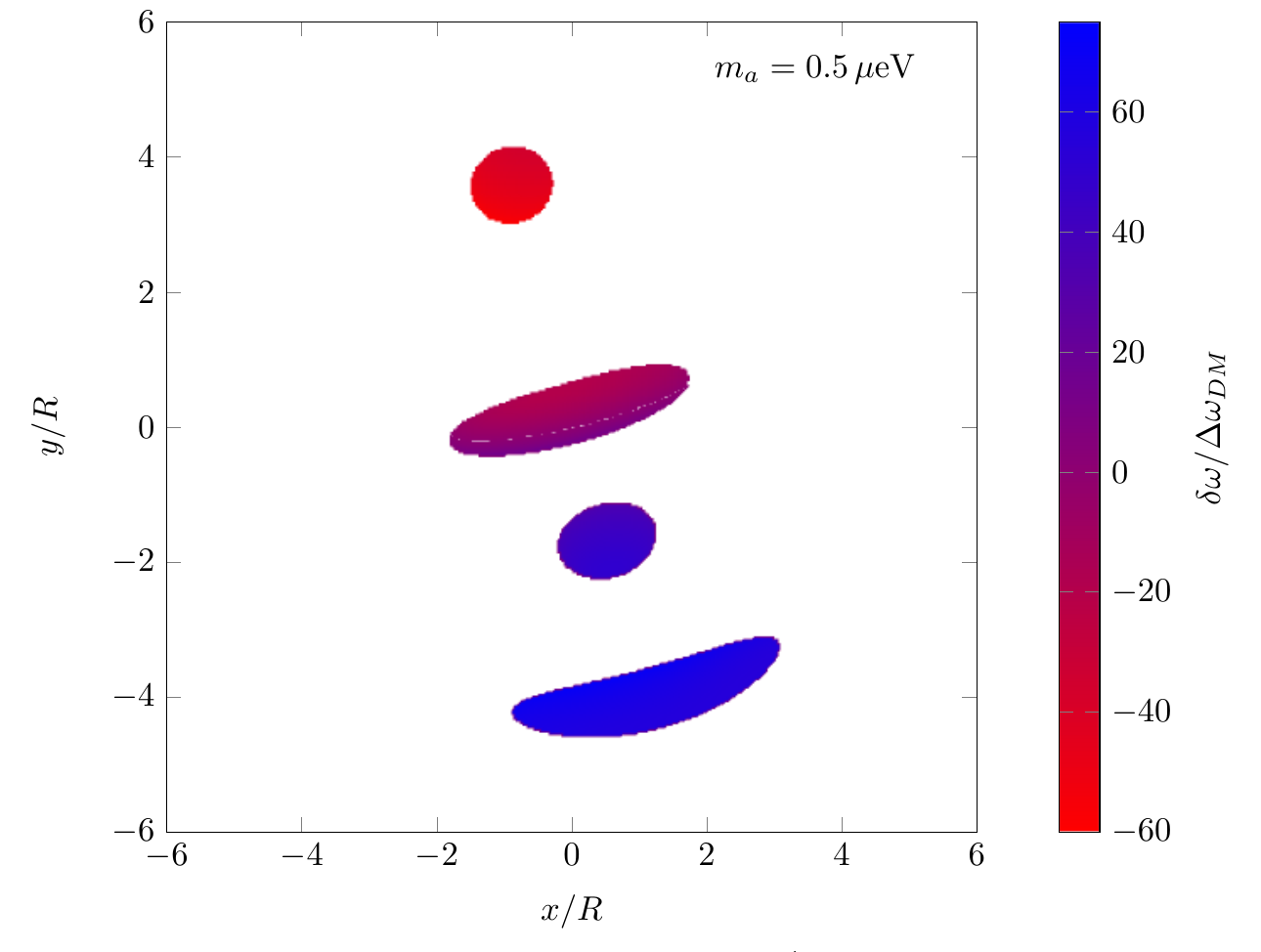}
    \caption{\textbf{Image of Doppler shift.} The red/blue shift in frequency as seen in the image plane at a single moment in time. The frequency shift was computed using eq.~\eqref{eq:DopplerEvolution}. The other model parameters are the same as in Fig.~\ref{fig:ImagePlaneComparison}. We show the frequency shift relative to that set by dark matter velocity dispersion: $\Delta \omega_{\rm DM} = \frac{1}{2} v_0^2  m_a$ with $v_0 = 6 \times 10^{-4}$ corresponding to a dark matter velocity $v_0 = 200 \,{\rm km }\,  {\rm sec^{-1}}$.}
    \label{fig:DopplerImage}
\end{figure}

One property of the signal which has remained less clear until now, is the precise width of the signal. We previously made an order of magnitude estimate for the Doppler broadening in \cite{Battye_2020}. Our treatment in this paper allows us to make the first systematic calculation of signal width and fully characterise the exact line shape of the signal. Understanding the signal width is important since the observation time required to achieve a given signal to noise ratio is given by the radiometer equation
\begin{equation}\label{Eqn:Radiometer}
t_{\rm int} = \left(\frac{2k_{\rm B}T_{\rm sys}}{A_{\rm eff}S_{\sigma}}\right)^2\frac{1}{\Delta \omega}\,,
\end{equation}
where $T_{\rm sys}$ is the system temperature, $S_{\sigma}$ is the flux density noise level, $A_{\rm eff}$ is the effective area of the telescope and $\Delta \omega$ is the signal width. We therefore see that determining the signal width is crucial in order to obtain the observation time needed to reach a given sensitivity in $g_{a \gamma \gamma}$.  In Fig.~\ref{fig:DopplerImage} we can see that photons emitted from different sections of the magnetosphere become blue or red-shifted according to eq.~\eqref{eq:DopplerEvolution}.

\begin{figure*}[t]   
    \centering
      \includegraphics[scale=0.75]{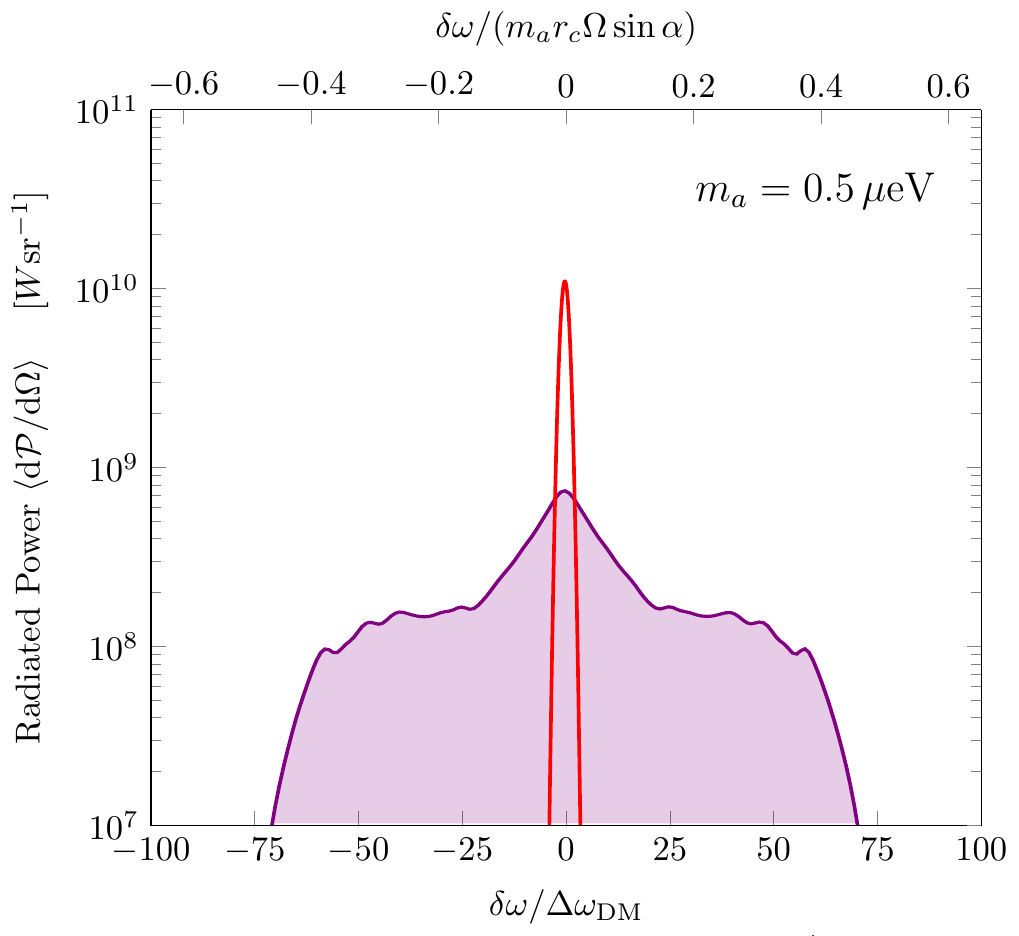}
        \includegraphics[scale=0.75]{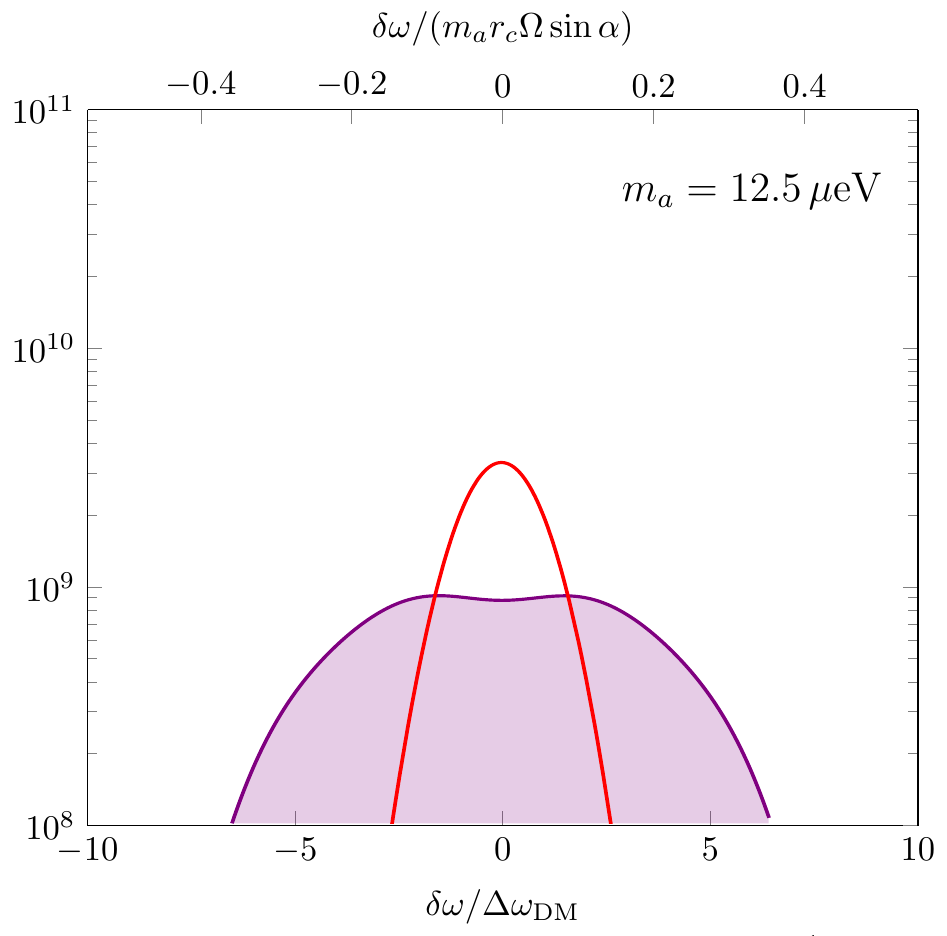}
    \caption{\textbf{Pulse-averaged line shapes.} Line profile (purple) of the Doppler broadened radio line computed via evolving eq.~\eqref{eq:DopplerEvolution} along rays. We took an observing angle $\theta = 36^\circ$. Other model parameters are as for Fig.~\ref{fig:PulseProfile}. The bottom axis shows the width relative to that set by velocity dispersion of dark matter $\Delta \omega_{\rm DM} = \frac{1}{2} v_0^2  m_a$ with $v_0 = 6 \times 10^{-4}$ corresponding to a dark matter velocity $v_0 = 200\, \rm{km} \, {\rm sec^{-1}}$. The top axis compares the width to the analytic estimate \eqref{eq:DopplerApprox}. We display the original un-broadened Gaussian line signal in red.}
     \label{fig:LineShapes}
\end{figure*}

\begin{figure*}
    \centering
    \includegraphics[scale=0.6]{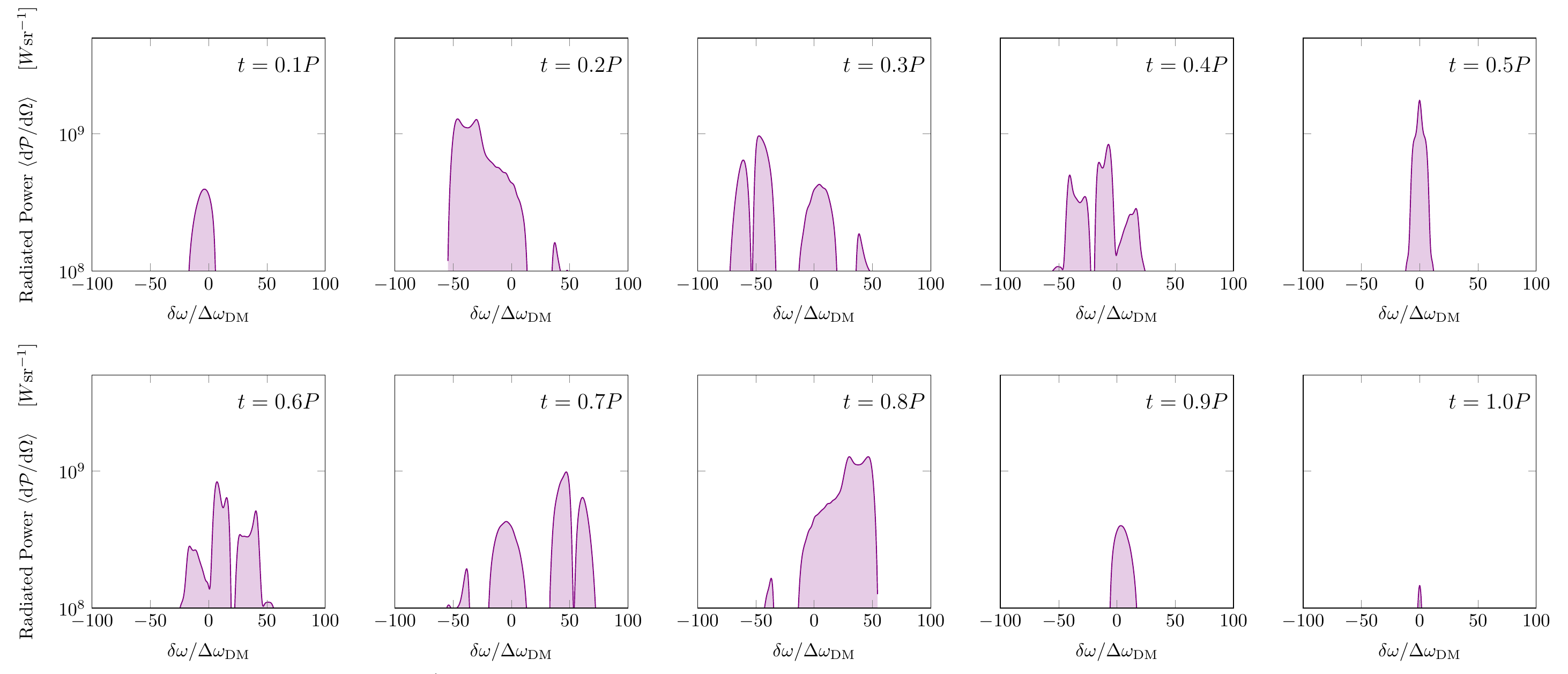}
    \caption{\textbf{Time-dependence of line shape.} We display the instantaneous line shape for $m_a = 0.5 \, \mu$eV at each point in the pulse period. Other values are as in fig.~\ref{fig:LineShapes}.}
    \label{fig:Instant_Line_Shape}
\end{figure*}

We plot the line shape of the signal for the maximum and minimum masses of interest in Fig.~\ref{fig:LineShapes}. We also display the instantaneous shape of the line signal in fig.~\ref{fig:Instant_Line_Shape}, from which we see that the Doppler broadening can be so sever that the unbroadened signal splits into several lines.These plots can be understood as follows. In both cases, the Doppler broadening is largest for those emission points which lie furthest from the centre of rotation, here the critical surface is moving with the greatest velocity and therefore imparts the most red/blue-shift (Fig.~\ref{fig:DopplerImage}). In addition, at lower masses, the critical surface lies farther from the centre of rotation, and so Doppler broadening decreases with increasing mass. 

\begin{figure*}[t]
    \centering
     \includegraphics[scale=0.4]{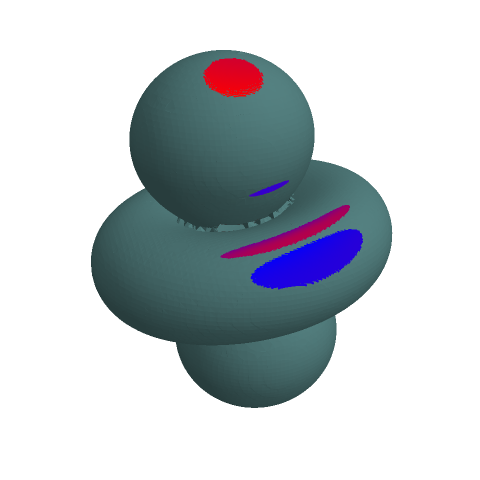}
     \hspace{2cm}
     \includegraphics[scale=0.4]{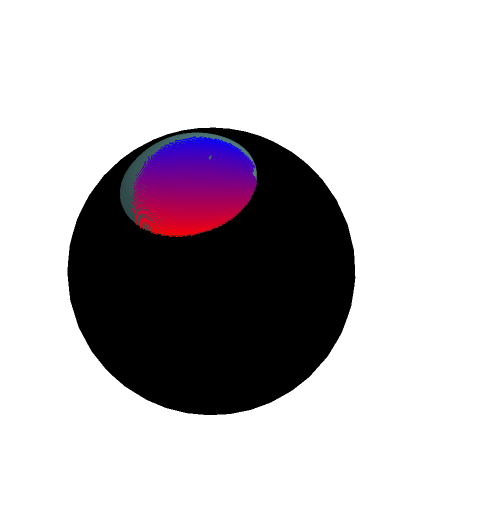}
    \caption{\textbf{Doppler shift back-projected onto surface emission points.} Projection of the redshifted/blue-shifted photons onto their emission points on the three-dimensional critical surface (gray) corresponding to the model parameters chosen in Fig.~\ref{fig:DopplerImage}. The left and right panels correspond to $m_a = 0.5 \mu$eV  and $m_a = 12.5 \mu$eV, respectively. For the higher mass, the critical surface only extends in the small polar cap regions above the stellar surface, which we illustrated with a black sphere.  Note the two images are not to the same scale with respect to one another.}
    \label{fig:SurfaceRedShift}
\end{figure*}

To understand the different line shapes in Fig.~\ref{fig:LineShapes} we note that for the lower mass $m_a = 0.5 \, \mu$eV, the emission points are reasonably democratically spread over the critical surface (Figs. \ref{fig:DopplerImage} and \ref{fig:SurfaceRedShift}) resulting in a more top-hat like profile, with a large amount of power coming from weakly Doppler broadened points near the equator, resulting in a central spike.  Meanwhile, for $m_a = 12.5\mu$eV the critical surface is sunk mostly within the star, with only small polar cap regions protruding from the surface of the star. The photons here consist of a single blue-shifted and red-shifted patch of equal size, resulting in two distinct bumps in the line in the right panel of Fig.~\ref{fig:LineShapes}. 


It is interesting to make an analytic estimate of the Doppler broadening.  Using the GJ density in eq.~\eqref{eq:GJDensity}, we can compute the frequency shift from eq.~\eqref{eq:DopplerBroadening1} to leading order in $\Omega$ which gives
\begin{align}\label{eq:dopplerbroadening}
&\delta \omega \simeq  \nonumber \\
&\left(\frac{4 \pi \alpha_{\rm EM}  R^3 \Omega B_0}{2 \, e \, m_{\rm e}} \right)\frac{3 \Omega \sin \alpha }{2 \omega_0} \,\int^\infty_0 dt \, \frac{\sin 2 \theta_0(t) \sin \phi_0(t) }{r^3_0(t)}\,,
\end{align}
where $\omega_0 \simeq m_a$ is the unperturbed frequency. The integral in general must be computed for each ray, and depends on the worldline $(r_0(t), \theta_0(t), \varphi_0(t))$. This is of course achieved in full by our numerical solutions. Nonetheless in order to make an order of magnitude estimate, we approximate a trajectory with $\theta_0(t)\simeq \theta_c = {\rm constant}$ and $\varphi_0(t) = \varphi_{c} = {\rm constant}$. We also approximate $dr/dt$ via the group velocity by taking $\omega_{p}^2(r) \simeq  m_a^2 (r_c/r)^3$ so that
\begin{eqnarray}
\frac{d r}{dt} \sim \sqrt{1 - \frac{\omega^2_{\rm p}(r)}{\omega_0^2}} ,
\end{eqnarray}
where $r_c$ is the characteristic radius of the critical surface which we define as
\begin{eqnarray}\label{eq:rcExp}
r_c  = m_a^{-2/3} \left(\frac{4 \pi \alpha_{\rm EM} \,R^3\,\Omega B_0}{2 \, e \, m_{\rm e}} \right)^{1/3}\,. 
\end{eqnarray}
This leads to 
\begin{equation}
\delta \omega \sim  \Omega m_a^2 r_c^3  \sin \chi  \sin  2 \theta_c \sin \phi_c  \,\int^\infty_{r_c} \, \frac{dr }{r^3 \sqrt{\omega_0^2 - \omega_{p}^2(r)}}\,.
\end{equation}
 Using $m_a \simeq \omega_0$ and performing the integral, we obtain the characteristic size of the Doppler broadening
\begin{eqnarray}\label{eq:DopplerApprox}
\frac{\delta \omega}{\omega_0} \sim r_c \Omega \sin \chi\,.
\end{eqnarray}
This reproduces the characteristic size of Doppler broadening based on our preliminary estimate in \cite{Battye_2020}. From \eqref{eq:rcExp}
we see that Doppler broadening grows with decreasing mass, where the radius of the critical surface is largest. From the upper axis in Fig.~\ref{fig:LineShapes}, we also see that eq.~\eqref{eq:DopplerApprox} gives a good order of magnitude estimate for the full numerical result.

Note the frequency distribution of the un-broadened signal is inherited from the Maxwellian dark matter velocity distribution $f_{\rm DM}(v) \propto e^{-v^2/v_0^2}$, so that with frequencies $\omega = m_a^2(1 + v^2)$ the power spectrum of the un-broadened signal is Gaussian:
\begin{equation}
 \mathcal{P}_0 (\omega) =\frac{\mathcal{P}_{\rm tot.}}{\sqrt{2 \pi \Delta \omega_{\rm DM}} } e^{- \frac{(\omega- ma)^2}{2 \Delta \omega_{\rm DM}}}\,,
\end{equation}
where $\Delta \omega_{\rm DM} = \frac{1}{2} v_0^2 m_a $ and $\mathcal{P}_{\rm tot.}$ is the total power, integrated across all frequencies. Each infinitesimal frequency band centered on $\omega'$ and contained within the original signal, there is a Doppler broadening 
\begin{eqnarray}
\delta(\omega - \omega' ) \rightarrow \mathcal{D}(\omega - \omega')\,,
\end{eqnarray}
where $\mathcal{D}(\omega - \omega')$ is also normalised to 1. Note this is a one-to-many map, which takes an infinitesimal part of the original signal $\mathcal{P}_0(\omega)$ and maps it to a broader set of frequency bins. By  linearity, we see that the original signal profile 
\begin{equation}
\mathcal{P}_{0} (\omega) =  \int d \omega'  \delta(\omega - \omega ') \mathcal{P}_0(\omega')\,,
\end{equation}
is mapped to a Doppler broadened signal 
\begin{equation}
 \mathcal{P}_{\rm Dopp.}(\omega)  = \int d \omega' \mathcal{D}(\omega  - \omega') \mathcal{P}_0(\omega')\,.
\end{equation}
In other words, the original signal becomes stretched by the Doppler broadening factor $\mathcal{D}(\omega - \omega')$, with the final result given by convolving the original line shape with the Doppler shift function $\mathcal{D}(\omega - \omega')$. The function $\mathcal{D}(\omega - \omega')$ is of course computed by tracing the frequency evolution of rays, and binning the power at each frequency, with the final line shapes shown in Fig.~\ref{fig:LineShapes}.



\section{Discussion}\label{sec:discussion}

In this work we have presented a framework in which to compute the detailed properties of radio lines resulting from the conversion of dark matter axions in the magnetospheres of pulsars. This reveals both an enhanced time-dependence relative to straight-line vacuum trajectories considered previously in \cite{Leroy:2019ghm} and allows us to rigorously compute both the shape and width of the signal in frequency space. 

We also provided arguments based on elementary optics to underpin our results.  We related the effective size of the image of the critical surface to the refractive index, which is a function both of the plasma density and gravity at the critical surface. This enables us to interpret the image size in terms of the competing lensing effects of gravity (attractive) and plasma (repulsive). 

 The Doppler broadening of the signal follows straight-forwardly from the geodesic equations for rays propagating through a medium with a time-dependent refractive index. We also derived an order of magnitude estimate for the width of the signal, which supports the estimate obtained in our previous work \cite{Battye_2020}.

Clearly what remains for future work is to provide a complete parameter scanning across axion masses, observing angles and magnetic alignment so as to produce the most conservative sensitivity curves for $g_{a \gamma \gamma}$. This could then be applied to astronomical data which would lead to the most robust constraints to date for the conversion of axion dark matter in neutron star magnetospheres.  


\begin{center}
\textbf{Note Added}
\end{center}
At the same time as our preprint for this paper was released, \cite{Witte:2021arp} also appeared which deals with similar questions addressed in this paper. Overall, both works reach the same broad conclusion that plasma-ray tracing is a necessary ingredient in accurately describing the signal, so that numerical ray-tracing should replace the naive modelling in \cite{ref:NS-Hook} in all future work. While our approach chooses a given observing direction and \textit{back-traces} rays onto the conversion surface (in the spirit of \cite{Leroy:2019ghm}), \cite{Witte:2021arp} uses a semi-stochastic approach, sampling over all-possible photon emissions directions at each point on the critical surface and  \textit{forward-propagating} rays to the sphere at infinity. For a large number of rays, this will then approach the true angular distribution of the signal. At this level, the difference is simply methodological and should not have and impact on observable results. 

The first difference between the two works is that \cite{Witte:2021arp} includes anisostropic effects in the dispersion relation due the magnetic field. Another notable physical difference between the two papers is that our work incorporates gravity into the ray-tracing equations via minimal coupling. We have subsequently added a supplementary appendix A to this paper to illustrate the effects of modifying the strength of gravity.  One can clearly see that the time-dependence of the signal is more pronounced as the effect of gravity is reduced, which seems to approach closer the results of \cite{Witte:2021arp} (no gravity) which reports greater time-variation of the signal than we do. This makes sense in light of the arguments of sec.~\ref{sec:grav}, where we describe how, owing to a larger impact parameter induced by gravity a given observer back-illuminates/samples a greater proportion of the emission surface. As a result, there is a more democratic spread of the power across the sky with increasing $r_s$, resulting in smoother angular/time variation of the signal. 

The authors of \cite{Witte:2021arp} also analyse many new effects at the level of the conversion probability itself which we do not include. Clearly in future work it would be interesting to combine all the novel effects considered in both references so as to incorporate the full array of corrections to previous work. 
One effect we note in passing, is the suggestion of de-phasing effects due to refraction causing the photon to go out of phase with the axion during the conversion process itself. Some preliminary estimates were made in \cite{Witte:2021arp} for the size of this effect. Clearly this is quite interesting, and the WKB estimates should be compared against full 3D simulations of the mixing equations (as hinted in \cite{Battye_2020}), perhaps along the lines of the comparison between \cite{McDonald:2020why,McDonald:2019wou} which compared WKB/ray-tracing against full electrodynamics simulations in a simple higher-dimensional geometry. 

It would also be extremely interesting to carry out detailed benchmarking of both codes against exactly solvable analytic examples, though such an analysis represents a significant undertaking and clearly lies beyond the scope of the present paper.

Finally, we remark that we have carried out convergence tests on our results with the ray integration step-size, number of rays and distance of the observing plane from the conversion surface.   During development, we also tested our code against simple spherically symmetric plasma distributions and their analytic trajectories presented in \cite{Rogers:2015dla}. We also tested our code in a toy-case for straight-line rays with a rotating ellipsoid with uniform surface luminosity, for which the pulse profile is simply given by the cross-sectional area of the ellipsoid.

\begin{center}
\textbf{Acknowledgements}
\end{center}

We are grateful to Ali Hashmi and Anthony Holloway for their assistance in cluster parallelisation of Mathematica. We also thank Sam Witte and the referee for helpful comments and discussions. JIM is supported by the F.R.S.-FNRS under the Excellence of Science (EOS) project No.~30820817 (be.h).   This work is also supported by the Collaborative Research Centre SFB 1258 of the Deutsche Forschungsgemeinschaft. SS is supported by a George Rigg Scholarship from the University of Manchester.

\appendix

\vspace{-1cm}
\section{Role of Gravity}\label{appendix:Gravity}

In this appendix we briefly display the effect of modifying the strength of gravity in the ray-tracing equations by varying the dimensionless ratio $r_{\rm s}/R$ in the ray tracing calculation. Note we keep the strength of gravity the same in determining the axion density via eq.~\eqref{eq:Liouville_Grav_Increase}, hence isolating the effect coming from the ray tracing.

\begin{figure}
    \centering
    \includegraphics[scale=1]{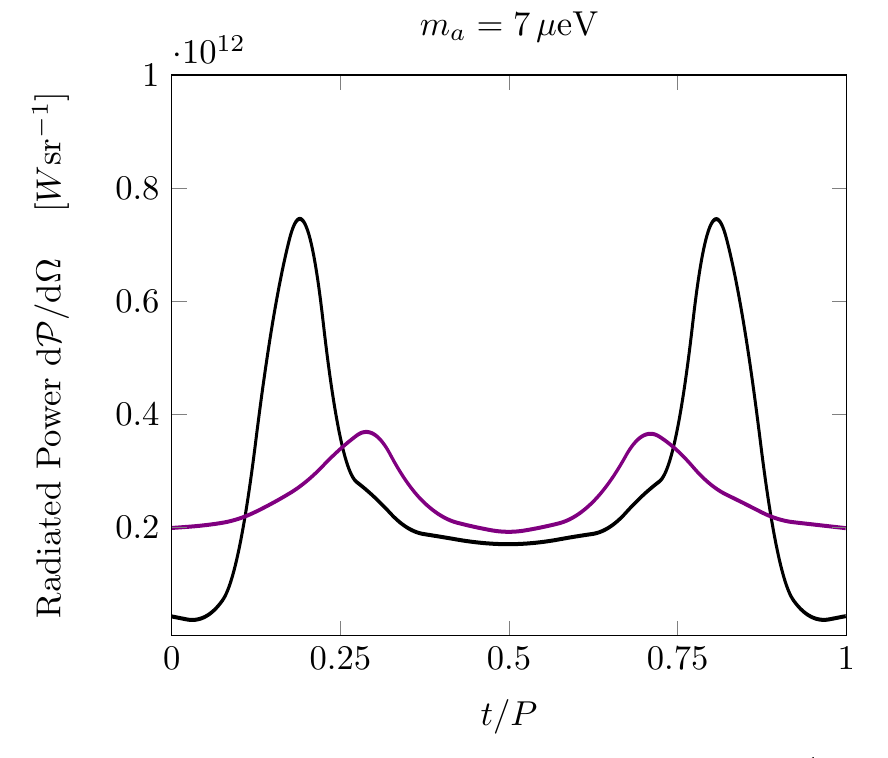}
    \caption{\textbf{Effect of Gravity on Pulse Profiles.} Pulse profiles with $r_s/R = 0.29$ (purple) and $r_s/R = 0.03 $ (black) for an axion mass $m_a = 7 \mu$eV and an observing angle $\theta = 36^\circ$. }
    \label{fig:Pulse_Profiles_Grav}
\end{figure}

In figs.~\ref{fig:Pulse_Profiles_Grav} we present the time dependent profile of frequency integrated power for two values $r_{\rm s}/R=0.29$, which is the standard value for the cases we have studied, and an artificially lower value of $r_{\rm s}/R=0.03$, around a factor 10 lower. We see that there is a much larger time variation in the case where the effect of gravity is lowered. In particular, the peaks in the profile are much larger. Qualitatively, the effect that we see is similar to that seen in \cite{Witte:2021arp} who report a much larger variation as a function of $t/P$.

\begin{figure*}
    \centering
    \includegraphics[scale=0.65]{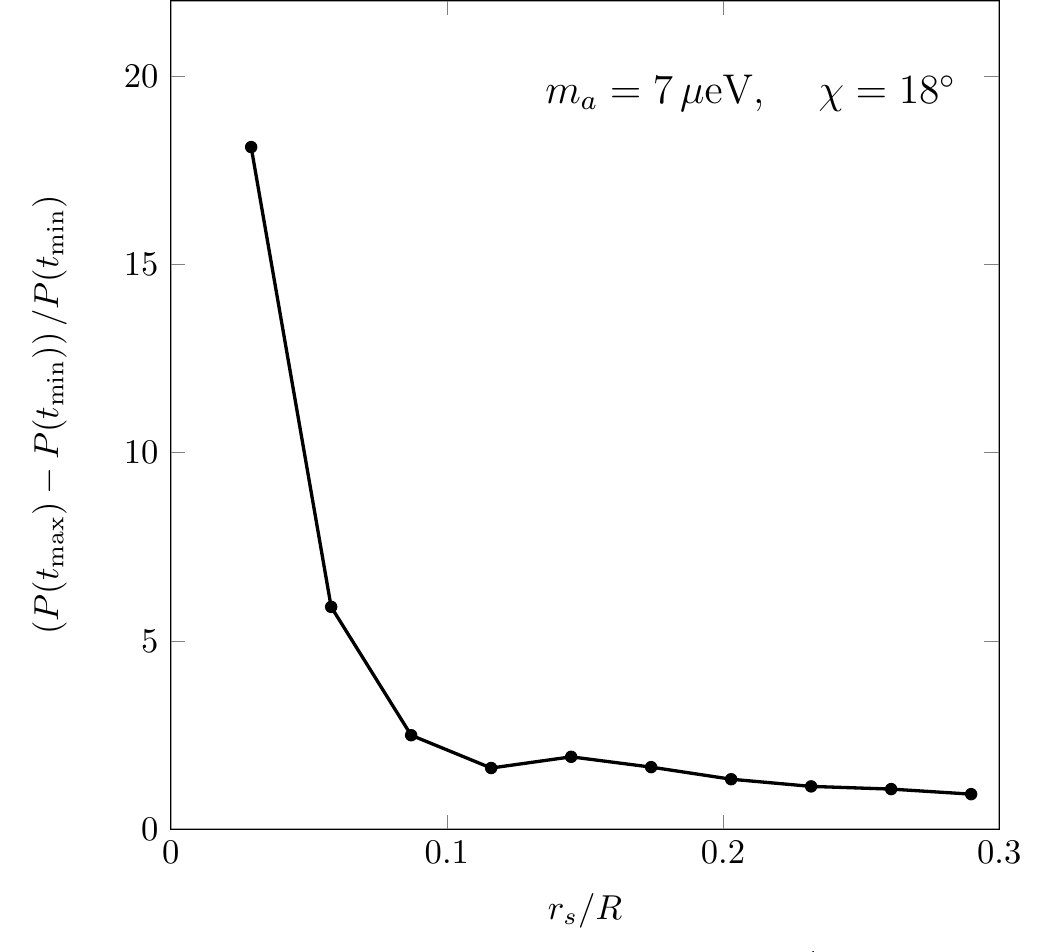}
    \includegraphics[scale=0.65]{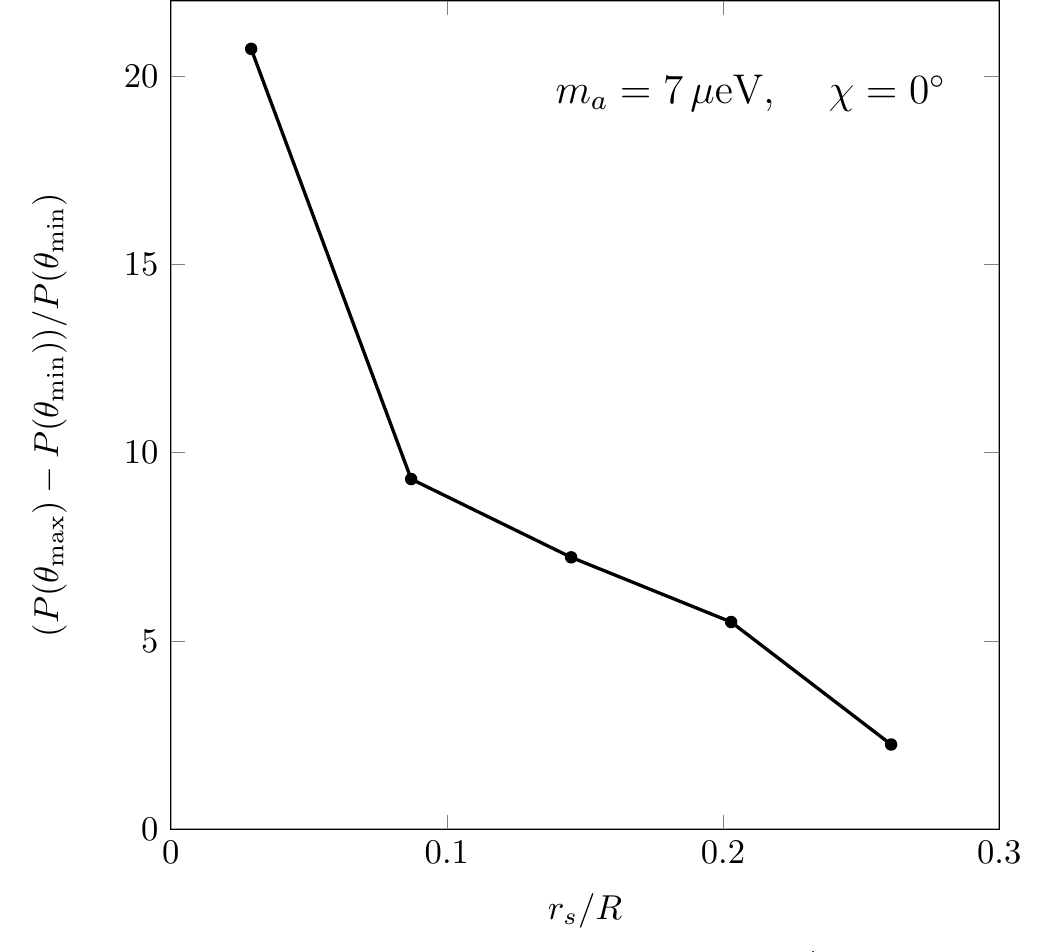}
    \caption{\textbf{Effect of Gravity on angular and time-variation.} We display the effect of gravity on the time-dependence of the signal (left panel) for an oblique rotator with $\chi = 18^\circ$ and the angular dependence (right panel) for an aligned rotator $\chi = 0$. We again took a value $m_a = 7 \mu$eV. Here $t_{\rm max/min}$ and $\theta_{\rm max/min}$ denote respectively the time and angle for which the power is maximum/minimum.}
    \label{fig:Gravity_Scaling}
\end{figure*}

To investigate this further we present two plots in \ref{fig:Gravity_Scaling}. The first is for an oblique rotator  with $\chi=18^{\circ}$ where we plot the fractional change in the power between the times when the power is at its maximum and minimum as in \ref{fig:Pulse_Profiles_Grav}. The second is for an aligned rotator with $\chi=0^{\circ}$, but this time varying the observation angle and plotting the fractional difference between in the power at its maximum and minimum. Both show a characteristic increase as $r_{\rm s}/R\rightarrow 0$ indicating that gravity is important. We note that the increase in the case of the aligned rotator can be estimated analytically. 

\begin{figure*}
    \centering
    \includegraphics[scale=0.65]{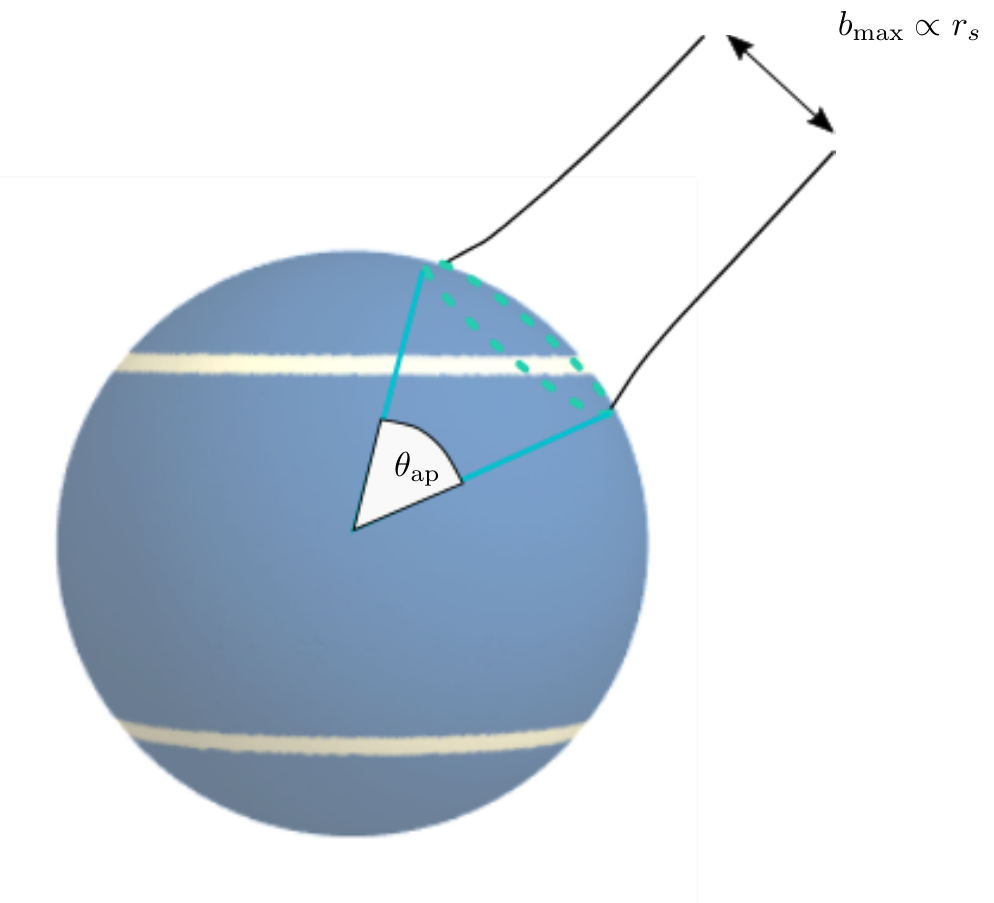}
    \includegraphics[scale=0.65]{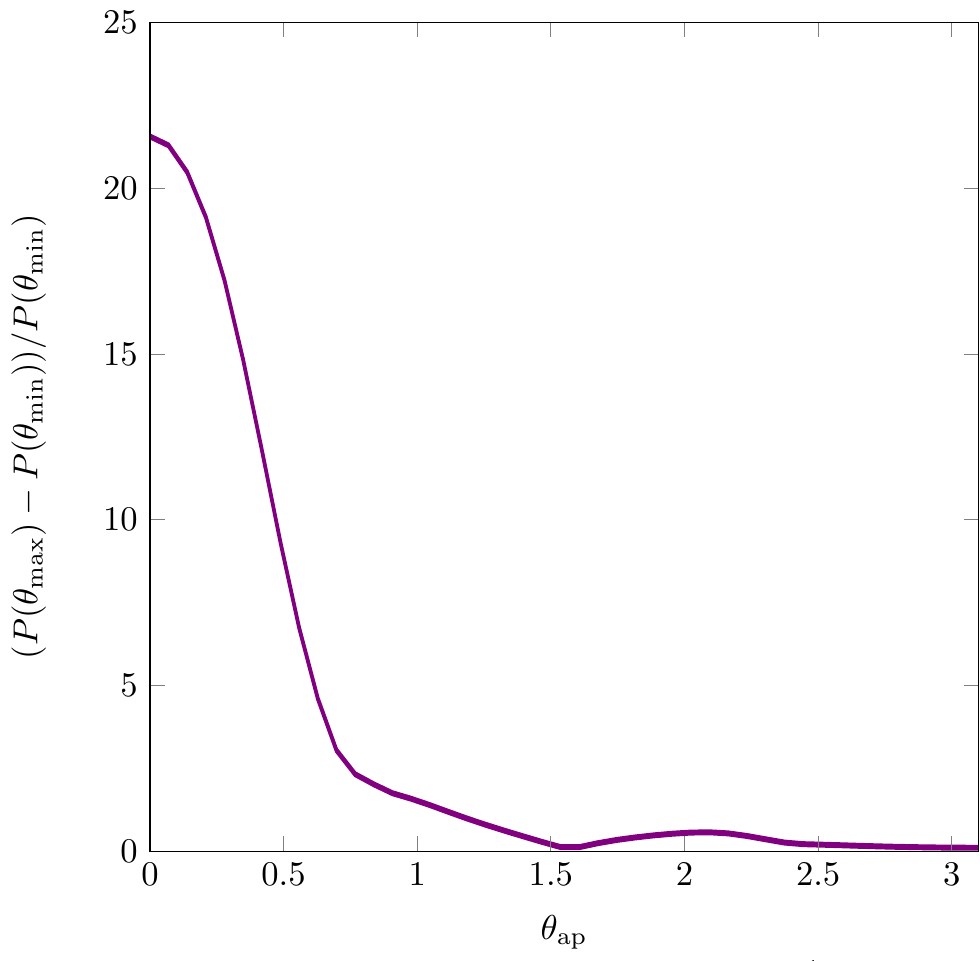}
    \caption{\textbf{Gravity in a toy analytic example.} We display the power coming from a circle on a spherical emission surface subtended by an angle $\theta_{\rm ap}$. We choose an emission surface (left panel) with two bright strips angular size $\Delta \theta = 0.01$ at positions $\theta = \pm \pi/4$. The right panel shows the resulting angular variation in the power with observing direction $\theta$ as a function of $\theta_{\rm ap}$. }
    \label{fig:toy_analytic}
\end{figure*}

In particular, we can understand this behaviour qualitatively in a simple analytic example where we consider the special case of a spherical plasma distribution discussed in sec.~\ref{sec:grav}. This has the added advantage that it circumvents any complicated ray-tracing codes and is based on simple physical reasoning. In this case the critical surface is simply a sphere radius $r_c$. The region illuminated in the image plane is a circle whose radius is described by some maximum impact parameter $b_{\rm max}$ of rays which strike the critical surface. These rays emanate from within a circle drawn on the critical surface (shown by the dashed circle in fig~\ref{fig:toy_analytic}). For simplicity, we take an isotropic surface emission. The luminosity is then given simply by integrating the surface emission contained within the dashed circle, which subtends an angle $\theta_{\rm ap}$ displayed in the figure. Hence, the larger the value of $r_c$, the larger the impact parameter $b_{\rm max}$, and hence the greater the size of the circle on the critical surface, and the larger the value of $\theta_{\rm ap}$. Therefore, varying $\theta_{\rm ap}$ is a proxy for varying $r_{\rm s}$. Of course the precise algebraic relation between $\theta_{\rm ap}$ an $r_{\rm s}$ depends on the form of the spherical plasma distribution on $r$, and can be determined by solving the geodesic equations for $\theta$ evolution in spherical polar coordinates as in \cite{Rogers:2015dla}, but for now we remain general and choose $\theta_{\rm ap}$ as a proxy scaling parameter for $r_{\rm s}$. This is sufficient to build up physical intuition for the results presented in this appendix. 

In the analytic example shown in the fig.~\ref{fig:toy_analytic}, we consider a critical surface which has two luminous strips angular size $\sim \Delta \theta$, with the surface luminosity being zero elsewhere. This is a proxy for the case of an aligned rotator in the GJ model where the strongest axion emission comes from similar strips above and below the equator near the line where $n_{\rm GJ} \simeq 0$ meets the stellar surface. 

The key conclusion is that this toy example produces precisely the same qualitative behaviour as in fig.~\ref{fig:Gravity_Scaling}. It helps us to understand that this scaling behaviour with $r_{\rm s}$ is a consequence of the fact that for a given observing direction, a larger proportion of the critical surface is probed, resulting in a more democratic spread of power across the sky. Overall, these examples suggest that gravity needs to be included in these calculations since it counteracts the effects of the plasma.

\section{Covariant radiative transport in plasmas}\label{appendix:RadiativeTransport}

We now derive the relation  \eqref{eq:IemeqlIobs} used in the main text. By definition, the phase space distribution for photons satisfies
\begin{eqnarray}\label{eq:phasespacedensity}
    \mathcal{F}(x^i, k^i) = \frac{dN}{d\mathcal{V}} = \frac{dN}{d^3\textbf{x}d^3\textbf{k}}\, .
\end{eqnarray}
where $dN$ is the number of photons in the phase space volume $d^3\textbf{x} d^3 \textbf{k}$. Formally, one must define phase space using a local coordinate system, since the background is inhomogeneous \cite{1975Hadrava}. The infinitesimal energy carried by photons frequency $\omega = \omega(\textbf{k})$ is then given by $dE_\omega = \omega dN$. Using the expression \eqref{eq:phasespacedensity} for $dN$, we arrive at
\begin{equation}\label{eq:dE}
    dE_{\omega} = \omega \mathcal{F}(x^i, p^i)d^3\textbf{k} \,d^3\textbf{x} \, ,
\end{equation}
We now re-write the momentum phase space element in terms of the refractive index 
\begin{equation}
n =k/\omega, \qquad k = \left| \textbf{k}\right|\,. 
\end{equation}
Firstly we change to spherical coordinates
\begin{equation}\label{eq:refractiveIndex}
d^3\textbf{k} = k^2 \, dk  \,  d\Omega(\hat{\textbf{k}})\,,
\end{equation}
where $d\Omega(\hat{\textbf{k}}) = \sin \theta_\textbf{k} d\theta_{\textbf{k}} d \varphi_{\textbf{k}}$ is the solid angle element in momentum space, defined with respect to some axis. Next we write $dk$ in terms of the refractive index by using 
\begin{eqnarray}\label{eq:dp}
    dk =d(n \omega) =  d \omega\left[ n + \omega \frac{d n}{d \omega}\right]\,.
\end{eqnarray}
The final term can be re-written in terms of the group velocity 
\begin{eqnarray}\label{eq:vg}
    v_{\rm g} = \left[ n + \omega \frac{d n}{d \omega}\right]^{-1}. 
\end{eqnarray}
Putting this together and combining eqs.~\eqref{eq:dE}-\eqref{eq:vg} we can re-write the momentum volume element as
\begin{equation}
d^3\textbf{k} = \frac{n^2\omega^2}{v_{\rm g}} \, d\omega \, d\Omega(\hat{\textbf{k}})\,.
\end{equation}
This leads immediately to 
\begin{equation}\label{eq:dEpolars}
    dE_{\omega} = \frac{\omega^3 n^2}{v_g} \mathcal{F}(x^i, k^i) \, d\omega \, d\Omega(\hat{\textbf{k}}) d^3 \textbf{x}\,.
\end{equation}
Finally, we invoke \textit{Liouville's theorem}, which states that by definition, the phase space density $\mathcal{F}$ is conserved along geodesics of Hamilton's equations. Hence we have
\begin{eqnarray}\label{eq:Loiuville}
    \frac{d}{d \lambda} \mathcal{F}\left( x(\lambda), k(\lambda)\right)= 0\,, 
\end{eqnarray}
where $\lambda$ is the wordline parameter and $x(\lambda), k(\lambda)$ satisfy eqs.~\eqref{eq:geodsic1} and \eqref{eq:geodesic2}.

To understand the power flow through a surface perpendicular to the wordlines, we introduce coordinates
\begin{equation}
 \textbf{x} = \left (x^1_{\perp}, x^2_\perp,x_{||} \right)\,,
\end{equation}
where  $x^1_{\perp},x^2_{\perp}$ lie in the plane perpendicular to the direction of propagation, corresponding to $x_\parallel$. The surface area element $dA = dx^1_{\perp}dx^2_{\perp}$.
If the group velocity along the worldline is $v_{\rm g}$, then we have $dx_{||} = v_{\rm g}dt$. In this coordinate system, the spatial volume element then becomes 
\begin{eqnarray}
 d^3\textbf{x} \rightarrow  v_{\rm g} dA dt.
\end{eqnarray}
Hence in these coordinates, eq.~\eqref{eq:dEpolars} reads
\begin{equation}
    dE_{\omega} = \frac{\omega^3 n^2}{v_g} \mathcal{F}(x^i, k^i) \, d\omega \, d\Omega(\hat{\textbf{k}})  d A dt\,.
\end{equation}
This allows one to connect the distribution function $\mathcal{F}$ to the Poynting flux which is the energy per unit time per unit area:
\begin{equation}\label{eq:Poynting}
   |\textbf{S}| =  \frac{dE}{dA dt}\, ,
\end{equation}
The specific intensity is then defined as
\begin{eqnarray}\label{eq:Intensity}
    I = \left| \frac{d\textbf{S}}{d\Omega d\omega} \right|\, ,
\end{eqnarray}
where $d\Omega$ gives the solid angle in the direction of propagation. 
On comparing eq.~\eqref{eq:dE} with \eqref{eq:Intensity}, one obtains that 
\begin{equation}
    \mathcal{F} = \frac{I}{n^2 \omega^3}\, ,
\end{equation}
and hence form eq.~\eqref{eq:Loiuville} conservation of  $\mathcal{F}$ along worldlines implies that 
\begin{eqnarray}
    \frac{I}{n^2 \omega^3} = \text{constant}\,,
\end{eqnarray}
along light rays. Hence we have the following relation between quantities at the point of emission on the critical surface, and points in the image plane:
\begin{eqnarray}
    \frac{I_{\rm obs}}{n^2_{\rm obs} \omega^3_{\rm obs}} = \frac{I_{\rm em}}{n^2_{\rm em} \omega^3_{\rm em}}\,,
\end{eqnarray}
where $\omega_{\rm obs}$ and $\omega_{\rm em}$ etc. are measured in fixed (i.e. accelerated rather than freely falling) frames.

\bibliographystyle{apsrev4-1}
\bibliography{Ref}

\end{document}